\documentclass[14pt]{article}
\usepackage[utf8]{inputenc}

\usepackage[margin=1.25in]{geometry}

\usepackage{hyperref}
\usepackage{enumitem}

\usepackage{amsthm}
\usepackage{amsmath}
\usepackage{tikz}

\newtheorem{theorem}{Theorem}
\newtheorem{corollary}[theorem]{Corollary}
\newtheorem{lemma}[theorem]{Lemma}

\newtheorem{definition}[theorem]{Definition}
\newtheorem{problem}[theorem]{Problem}

\newtheorem*{claim}{Claim}
\theoremstyle{remark}
\newtheorem*{claimproof}{Proof of claim}

\usepackage{amssymb}

\title{On classes of bounded tree rank, their interpretations, and efficient sparsification} 

\author{
  Jakub Gajarsk\'y\footnote{University of Warsaw (\texttt{gajarsky@mimuw.edu.pl}). This author is supported by the Polish National Science Centre SONATA-18 grant number 2022/47/D/ST6/03421. Parts of this work were developed while this author received funding from the European Research Council (ERC), grant agreement No
948057 — BOBR.} 
  \and
  Rose McCarty\footnote{Georgia Institute of Technology (\texttt{rmccarty3@gatech.edu}). This author is supported by the National Science Foundation under Grant No. DMS-2202961.} 
}

\date{}

\renewcommand{\phi}{\varphi}

\usepackage{todonotes}

\newcommand{\N}{\mathbb{N}}
\newcommand{\C}{\mathcal{C}}
\newcommand{\D}{\mathcal{D}}
\newcommand{\F}{\mathcal{F}}
\newcommand{\wh}{\widehat}

\newcommand{\SSS}{\mathcal{S}}

\newcommand{\dist}{\mathrm{dist}}

\newcommand{\scol}{\mathrm{scol}}
\newcommand{\adm}{\mathrm{adm}}

\usepackage{textpos}

\begin{document}

\maketitle

 \begin{textblock}{20}(-1.9, 8.2)
  \includegraphics[width=35px]{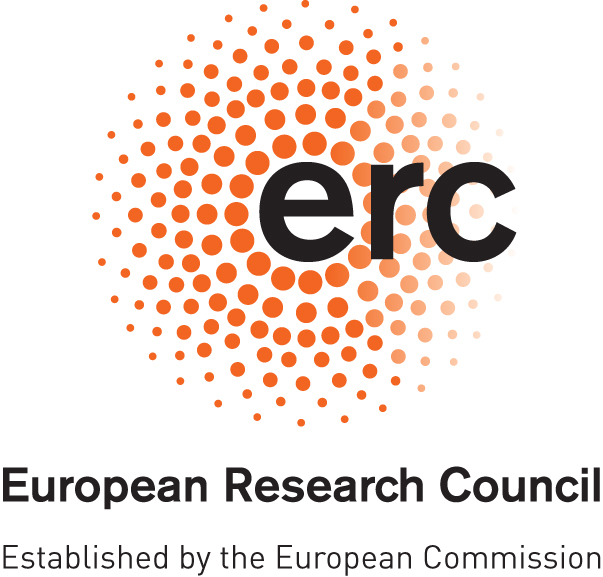}%
 \end{textblock}
 \begin{textblock}{20}(-2.15, 8.5)
  \includegraphics[width=50px]{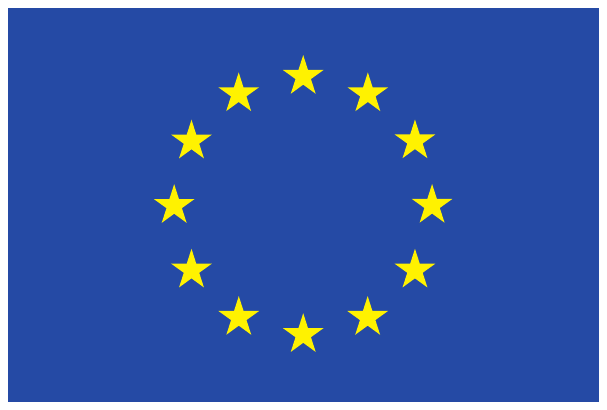}%
 \end{textblock}

\begin{abstract}
Graph classes of bounded tree rank were introduced recently in the context of the model checking problem for first-order logic of graphs. These graph classes are a common generalization of graph classes of bounded degree and bounded treedepth, and they are a special case of graph classes of bounded expansion. We introduce a notion of decomposition for these classes and show that these decompositions can be efficiently computed.
Also, a natural extension of our decomposition leads to a new characterization and decomposition for graph classes of bounded expansion (and an efficient algorithm computing this decomposition).

We then focus on interpretations of graph classes of bounded tree rank. We give a characterization of graph classes interpretable in graph classes of tree rank $2$. Importantly, our characterization leads to an efficient sparsification procedure: For any graph class $\C$ interpretable in a graph class of tree rank at most $2$, there is a polynomial time algorithm that to any $G \in \C$ computes a (sparse) graph $H$ from a fixed graph class of tree rank at most $2$ such that $G = I(H)$ for a fixed interpretation $I$. To the best of our knowledge, this is the first efficient ``interpretation reversal'' result that generalizes the result of Gajarsk\'y et al. [LICS 2016], who showed an analogous result for graph classes interpretable in classes of graphs of bounded degree.
\end{abstract}

\section{Introduction}
\label{sec:intro}
The graph classes and problems studied in this paper are motivated by considering the  first-order (FO) model checking problem for graphs.
This problem asks, given a (finite) graph $G$ and sentence $\phi$ as input, whether $G$ is a model of $\phi$. This problem is known to be PSPACE-hard, and so we do not expect to obtain a polynomial algorithm solving it. This has motivated the study of the FO model checking problem from the perspective of parameterized complexity, which has led to the discovery of many beautiful connections between structural and algorithmic graph theory and (finite) model theory. 

In the parameterized setting (where we consider the size of the formula $\phi$ as the parameter), we can easily obtain a brute-force algorithm with runtime $n^{O(|\phi|)}$, the so-called XP algorithm.
However, we are interested in the existence of algorithms with runtime $f(|\phi|)\cdot n^c$, where $c$ is some fixed constant (FPT algorithms). We do not expect to obtain such an algorithm if the input graphs come from the class of all graphs, since this problem is AW[$\ast$]-complete.
However, for various structurally restricted graph classes we know that such an algorithm exists. 
Identifying these graph classes is the topic of a long line of research, and recently a lot of progress has been made towards understanding the boundaries of efficient tractability~\cite{tww4,dreier2023firstorder, dms}.


For graph classes that admit a model checking algorithm with runtime $f(|\phi|)\cdot n^c$ one can ask what is the dependence of the runtime on the size of the formula $\phi$. Unfortunately, for most graph classes this dependence is \emph{non-elementary}, that is, $f(k)$ grows like a tower of twos whose height grows with $k$.
By a result of Frick and Grohe~\cite{frick2004complexity}, we know that this cannot be avoided even when $\C$ is the class of all trees. 
While this result may seem very limiting, it turns out that the landscape of graph classes that admit an elementary model checking algorithm is surprisingly rich.
Classical examples of such graph classes include classes of graphs of bounded degree~\cite{frick2004complexity}, bounded treedepth, and bounded shrubdepth~\cite{gajarsky-hlineny}. (The last two results were obtained in the more general context of MSO model checking.)
The interest in this problem was renewed recently when Lampis~\cite{lampis} established elementary model checking for graph classes of bounded pathwidth. After this, Gajarsk\'y, Pilipczuk, Sokołowski, Stamoulis and Toruńczyk~\cite{gajarsky2024elementary} introduced graph classes of bounded tree rank and proved that these classes also admit (under certain restrictions, see the fourth bullet point below) an elementary FO model checking algorithm. 

Classes of bounded tree rank, introduced in~\cite{gajarsky2024elementary}, can be defined as follows (in the definition, the \textit{depth} of a tree $T$ is the number of edges on a shortest leaf-to-root path in $T$).
\begin{definition}
\label{def:tree-rank}
A graph class $\C$ has \emph{tree rank} at most $d$ if for every $r \in \mathbb{N}$ there exists a tree $T$ of depth $d$ such that no $G \in \C$ contains $T$ as an $r$-shallow topological minor\footnote{We note that this definition of tree rank is seemingly off by $1$  from the original definition in~\cite{gajarsky2024elementary}. However, we measure the depth of trees differently than in~\cite{gajarsky2024elementary}, and so the two definitions ultimately coincide.}.  
\end{definition}

\noindent For example, one can easily see that graph classes of bounded degree have tree rank at most $1$, that the class of all trees of height $d$ has tree rank $d$, and that the class of all trees does not have bounded tree rank.

We briefly summarize some of the basic properties of graph classes of bounded tree rank established in~\cite{gajarsky2024elementary}:
\begin{itemize}
    \item They generalize graph classes of bounded degree, bounded treedepth, and bounded pathwidth.
    \item They are strictly less general than classes of graphs of bounded expansion.
    \item The tree rank of a graph class $\C$ is at most $d$ if and only if Splitter has a winning strategy in the so-called ``$d$-round batched Splitter game'' for every $G \in \C$. This game is a natural variant of the Splitter game, which characterizes nowhere dense graph classes and was introduced by Grohe, Kreutzer, and Siebertz in their landmark paper~\cite{gks}.
    \item If $\C$ is a class of bounded tree rank and there exists an elementary function $f: \N \to \N$ such that the size of $T$ in Definition~\ref{def:tree-rank} is bounded by $f(r)$, then there is an elementary FO model checking algorithm for $\C$.
    \item If $\C$ is a monotone graph class of unbounded tree rank, then $\C$ has no FO model checking algorithm with elementary dependence on the size of the formula unless FPT=AW[$\ast$].
\end{itemize}

We note that the concepts considered above (bounded degree, bounded treedepth, bounded tree rank, bounded expansion, and nowhere denseness) are all examples of classes of \emph{sparse} graphs.
Given how naturally the concept of bounded tree rank fits into the general theory of sparse graphs introduced by Ne\v{s}et\v{r}il and Ossona de Mendez~\cite{nevsetvril2012sparsity}, we suspect that it will play an important role in structural and algorithmic graph theory, and that further investigation of its structural and combinatorial properties is desirable. 

Since classes of sparse graphs are mostly well-understood (one exception being classes of bounded tree rank), in recent years there has been a trend to study more general classes that can be obtained from classes of sparse graphs by means of \emph{interpretations} or \emph{transductions}, which are graph transformations based in logic.
This point of view is also very relevant for classes of bounded tree rank, but we first discuss it in the general setting of sparse graphs. We focus on the simpler setting of interpretations. Essentially, an interpretation $I$ is given by a formula $\psi(x,y)$ (we will actually use a slightly more complicated setting; see Section \ref{sec:prelims}). When applied to a graph $H$, the result is a new graph $I(H)$ with the same vertex set as $H$ and with edge set $\{uv~:~H \models \psi(u,v)\}$. This notion generalises easily to graph classes by setting $I(\C) = \{I(H)~:~H \in \C\}$. Finally, we say that a graph class $\D$ is \emph{interpretable} in a graph class $\C$ if there exists an interpretation $I$ such that  $\D \subseteq I(\C)$.

As mentioned above, the recent trend is to study graph classes interpretable in sparse (or nowhere dense) classes of graphs. In particular, the model checking problem has been considered extensively, and was recently fully solved for such graph classes.



\begin{theorem}[\cite{dms}]
\label{thm:MC}
Let $\C$ be a graph class interpretable in a nowhere dense graph class. Then there exists an FPT model checking algorithm for $\C$.
\end{theorem}

In the context of graph classes interpretable in classes of sparse graphs, it is also natural consider the following problem, which we refer to as the ``efficient sparsification problem'' or ``interpretation reversal problem''. This problem was considered in~\cite{bd_deg_interp2} in the context of graph classes interpretable in classes of graphs of bounded degree.


\begin{problem}
\label{prob:sparsification}
Let $\C$ be a graph class interpretable in a nowhere dense graph class. Show that there exists a nowhere dense graph class $\D$, an interpretation $I$, and a polynomial time algorithm that given $G \in \C$ as input computes a graph $H \in \D$ such that $G = I(H)$.
\end{problem}
It is well-known (and easy to argue, see for instance~\cite{bd_deg_interp2}) that solving Problem~\ref{prob:sparsification} would lead to an alternative proof of Theorem~\ref{thm:MC}. Indeed, before the result of~\cite{dms} was established, this was considered the main line of attack on the model checking problem on graph class interpretable in a nowhere dense graph class. However,
Problem~\ref{prob:sparsification} turned out to be very challenging, and
despite considerable effort essentially no success has been achieved in solving it. Instead, Theorem~\ref{thm:MC} was proved by adapting the techniques for classes of sparse graphs from~\cite{gks} to the dense setting. Despite this, Problem~\ref{prob:sparsification} remains of considerable interest, and solving it would likely lead to many new insights on the structure of graph classes interpretable in classes of sparse graphs. 

Coming back to classes of bounded tree rank, it is natural to try to obtain an elementary analogue of Theorem~\ref{thm:MC}.
\begin{problem}
\label{prob:MC2}
Let $\C$ be a graph class interpretable in a graph class of bounded tree rank. Show that there exists an elementary FPT model checking algorithm for $\C$.
\end{problem}
In light of the results of~\cite{gajarsky2024elementary}, it is possible that one needs to put some extra restriction on $\C$ -- perhaps requiring that the size of trees avoided as $r$-shallow topological minors is bounded by an elementary function of $r$. However, it is currently unknown whether this is necessary, even for the original theorem from~\cite{gajarsky2024elementary}. 

In relation to interpretations of graph classes of bounded tree rank, the authors of~\cite{gajarsky2024elementary} introduced the more general graph classes of bounded \emph{rank}, and conjectured that these are precisely the interpretations of colored graph classes of bounded tree rank. In order to attack this conjecture, or Problem~\ref{prob:MC2}, it is natural to consider the following variant of Problem~\ref{prob:sparsification}.
\begin{problem}
\label{prob:sparsification2}
Let $\C$ be a graph class interpretable in a graph class of bounded tree rank. Show that there exists a graph class $\D$ of bounded tree rank, an interpretation $I$, and a polynomial time algorithm that given $G \in \C$ as input computes a graph $H \in \D$ such that $G = I(H)$.
\end{problem}

\paragraph*{Our contribution}

For sparse graphs we introduce the notion of the \emph{$(r,m)$-rank} of a vertex. Roughly speaking, the $(r,m)$-rank of a vertex $v \in V(G)$ is a positive integer that measures how complicated the $r$-neighborhood of $v$ is with respect to the parameter $m$. We also allow the $(r,m)$-rank to be $\infty$ if the $r$-neighborhood of $v$ is too complicated. Then the \emph{$(r,m)$-ranking} of a graph $G$ is the function $f: V(G) \to \N \cup \{\infty\}$ that assigns to each vertex of $G$ its $(r,m)$-rank. This definition has the advantage of being much more localized than Definition~\ref{def:tree-rank} of tree rank. Yet we prove that it can also be used to define tree rank, as follows.

\begin{theorem}
\label{thm:rankings=treerank}
For any $d \in \N$ and any graph class $\C$, the following are equivalent:
\begin{itemize}
    \item The tree-rank of $\C$ is at most $d$.
    \item For every $r \in \mathbb{N}$ there exists $m \in \mathbb{N}$ such that every vertex of every graph in $\C$ has $(r,m)$-rank at most~$d$.
\end{itemize}

\end{theorem}

Vertex rankings provide a natural notion of decomposition for graph classes of bounded tree rank. We will show that this decomposition can be computed in elementary FPT runtime with respect to the parameters $r$ and $m$ (see Theorem~\ref{thm:ranking-compute}). Moreover, the notion of $(r,m)$-ranking (and the FPT algorithm computing it) very naturally extends to graph classes of bounded expansion.

\begin{theorem}
\label{thm:rankings=be}
The following are equivalent for any graph glass $\C$:
\begin{itemize}
    \item The class $\C$ is of bounded expansion.
    \item For every $r \in \N$ there exists $m\in \N$ such that every vertex of every graph in $\C$ has finite $(r,m)$-rank, that is, has $(r,m)$-rank not equal to $\infty$.
\end{itemize}
\end{theorem}

\noindent We note that the $(r,m)$-rank of a vertex is preserved under graph automorphisms. This fact that rankings are ``canonical'' is one of the key advantages of Theorem~\ref{thm:rankings=be}.

One of the main motivations for proving alternate characterizations of sparse graph classes (like Theorems~\ref{thm:rankings=treerank} and~\ref{thm:rankings=be}) is the efficient sparsification problems discussed earlier (Problems~\ref{prob:sparsification} and~\ref{prob:sparsification2}). In particular, we hope that this definition of $(r,m)$-rank can be generalized to accommodate for interpretations of classes of bounded expansion. While we cannot yet take care of this more general case, we are able to use the insights obtained from $(r,m)$-rank to find the following structural characterization of interpretations of classes of tree rank~2. 

\begin{theorem}
\label{thm:characterization}
The following are equivalent for any graph glass $\C$:
\begin{itemize}
    \item The class $\C$ is interpretable in a graph class of tree rank $2$.
    \item The class $\C$ is a perturbation of a locally almost near-covered graph class.
\end{itemize}
\end{theorem}
We defer the definition of locally almost near-covered graph classes to Section~\ref{sec:intereps_bd_trr}.
Our characterization leads to the following algorithmic sparsification result. For technical reasons we restrict ourselves to classes of graphs interpretable in classes of tree rank $2$ for which the bounds in the definition of tree rank can be efficiently computed. We call such classes \emph{efficiently bounded} (see Section~\ref{sec:tr2}) for precise definition.

\begin{theorem}
\label{thm:sparsify_alg}
Let $\C$ be a graph class interpretable in an efficiently bounded graph class of tree rank $2$. Then there exists a graph class $\D$ of tree rank $2$, an interpretation $I$, and a polynomial time algorithm that given $G \in \C$ as input computes a graph $H \in \D$ such that $G = I(H)$.
\end{theorem} 

We remark that the degree of the polynomial in Theorem~\ref{thm:sparsify_alg} depends on the class $\C$.

While at first sight Theorem~\ref{thm:sparsify_alg} might seem like a modest improvement on the analogous result of~\cite{bd_deg_interp2} for graph classes of bounded degree (or equivalently, classes of tree rank at most $1$), this is the first progress on the challenging Problem~\ref{prob:sparsification} since its introduction that applies to graph classes of unbounded treewidth. (For graph classes of bounded treewidth such a result is given in~\cite{NesetrilRMS20}.)
Moreover, our proof 
introduces new techniques and ideas that
may be useful for solving the sparsification problem (Problem~\ref{prob:sparsification}) in greater generality. 

In particular, as part of the proof of Theorem~\ref{thm:characterization} we prove a lemma (Lemma~\ref{lem:Rose_lemma}) about the behaviour of the $k$-near-twin relation on graphs that do not contain a half-graph as a semi-induced subgraph. This lemma may be of independent interest in the context of (monadically) stable graph classes, which play a prominent role in recent developments establishing connections between (finite) model theory and algorithmic graph theory~\cite{dreier2023firstorder, indiscernibles2023}.




\paragraph*{Related work}

In~\cite{gajarsky2024elementary} it was shown that classes of tree rank at most $d$ can be characterized using the \emph{batched Splitter game} with $d$ rounds. Roughly speaking, in this game two players take turns, one of them trying to simplify the graph, and the other trying to keep the graph as complicated as possible. The result of~\cite{gajarsky2024elementary} states that the player trying to simplify the graph (this player is called Splitter) wins in at most $d$ rounds.

While this characterization is very nice and useful, it is a characterization in terms of a dynamic process, and as such it does not directly provide us with a useful notion of decomposition (in the sense of giving us a concrete compact object on which one can design algorithms). One could of course consider using as a decomposition the game tree arising from a winning play by Splitter, but this tree has size of order $n^d$. It is known that this can be circumvented by combining the Splitter game with \emph{sparse neighborhood covers} as introduced in~\cite{gks}, but working with the resulting object is technically demanding. Compared to this, vertex rankings give us a static decomposition on which one can use bottom-up inductive arguments and can design algorithms.

For graph classes of bounded expansion, vertex rankings of a graph (which certify that it has only vertices of finite rank) are closely related to strong coloring orders~\cite{zhu2009colouring} and admissibility orders~\cite{dvovrak2013constant}. These are total orders on the vertex set of a graph that satisfy certain properties, and which have proven to be very useful for showing properties of graphs of bounded expansion. The main idea used in the definition of rankings, when one checks whether a removal of a small set of vertices can separate a vertex $v$ from the set of previously processed vertices, has been used before (see Appendix A1 in~\cite{flipwidth}). This was again in the context of defining a suitable total order on the vertex set of a graph. Compared to total orders, our definition of rankings essentially leads to a pre-order on the vertex set of a graph, and therefore it can capture situations when two vertices are equally complicated (have the same rank).

Regarding the sparsification problem, in~\cite{bd_deg_interp2} it was shown that there is a polynomial time algorithm for sparsifying graphs from graph classes that are interpretable in classes of bounded degree. 
Another result related to efficient sparsification is~\cite{mapgraphs}, where the authors showed how to efficiently interpret the class of map graphs in a nowhere dense class of graphs. Finally, the already mentioned result~\cite{NesetrilRMS20} establishes a sparsification result for graph classes interpretable in graph classes of bounded treewidth, extending earlier results in~\cite{NesetrilRMS20linear}. 

\paragraph*{Outline of our approach}
We now briefly outline the approach used to prove  Theorems~\ref{thm:characterization} and~\ref{thm:sparsify_alg}. This approach builds on the ideas used 
in~\cite{bd_deg_interp2} to sparsify classes of graphs interpretable in graph classes of bounded degree. 
The key notion behind this result was that of \emph{$k$-near-twin} vertices. We say that two vertices $u,v$ of $G$ are $k$-near-twins if $|N^G(v) \Delta N^G(u)| \le k$, i.e. if they have the same neighborhoods with at most $k$ exceptions. The idea behind the proof given  in~\cite{bd_deg_interp2} is that graphs from graph classes interpretable in classes of bounded degree have a simple structure -- for such graphs we can find a small $k$ such that the $k$-near-twin relation is an equivalence on $V(G)$ with a bounded number of classes. This is then easily used to find a bounded number of \emph{flips} (edge complementations between two subsets of $V(G)$) to produce a sparse graph $H$ from which $G$ can be recovered by an interpretation.

In the case of classes of graphs interpretable in graph classes of tree rank $2$, the structure of the $k$-near-twin relation is much more complicated. 
In what follows we will actually focus on analysing this relation on graphs from graph classes interpreted in classes of tree rank 2 by an \emph{interpretation of bounded range}. 
This is an interpretation in which the formula $\psi(x,y)$ has the property that there is a number $b$ such that for every $G$ and $u,v \in V(G)$ we have that if $\dist_G(u,v) > b$, then $G \not\models \psi(u,v)$. In other words, such interpretation will never create edges between vertices that are far apart in $G$. 
The case of interpretations of bounded range forms the techincal core of our approach, since the reduction from the general case to the case of bounded range can be achieved by using existing tools (see Section~\ref{sec:full_interps}).
For the rest of this overview, let us fix a graph class $\C$  interpretable in a class of graphs of tree rank at most $2$ by an interpretation of bounded range. 

We now proceed with analysing the $k$-near-twin relation on a graph $G$ from $\C$.
It is useful to think of this relation in terms of the 
\emph{$k$-near-twin graph} of $G$, denoted by $NT_k(G)$. This graph has the same vertex set as $G$, and two vertices are adjacent in $NT_k(G)$ if they are $k$-near-twins in $G$. In the case of interpretations of bounded degree, this graph was a collection of a small number of cliques. In our case of interpretations of classes of tree rank $2$, the connected components of the graph $NT_k(G)$ are not cliques, and there can be arbitrarily many of them. Moreover, the connected components of $NT_k(G)$ can have arbitrarily large diameter -- this is important because if we could find a bound $d$ such that the diameter of all connected components of $NT_k(G)$ was at most $d$, then all vertices in any component $C$ would be $kd$-near-twins, and this could be exploited for sparsification.
In our proofs of Theorems~\ref{thm:characterization} and~\ref{thm:sparsify_alg} we overcome all these difficulties. The key insight (Lemma~\ref{lem:Rose_lemma}) is that even though the connected components of $NT_k(G)$ can have arbitrarily large diameter, we can nevertheless guarantee that any two vertices in the same connected component are $k'$-near-twins, for $k'$ depending only on $k$ and the order of largest half-graph in $G$. Using this, we sparsify  $G$ as follows: For a suitably chosen $k$, we consider $NT_k(G)$ and create a partition $\F$ of $V(G)$ by putting two vertices in the same part if they are in the same connected component of $NT_k(G)$. The aforementioned Lemma~\ref{lem:Rose_lemma} then guarantees that the vertices in the same part $A$ of $\F$ are pairwise $k'$-near-twins. 
We then create a sparse graph $\SSS(G)$ from $G$ as follows: If $A$ and $B$ are two large parts of $\F$ such that there are almost all edges between $A$ and $B$ (see the proof of Lemma~\ref{lem:main} for precise meaning of `large' and `almost all'), we complement the adjacency between them, create new vertices $v_A$ and $v_B$ adjacent to all vertices of $A$ and $B$, respectively, and create and edge between $v_A$ and $v_B$. 
The introduction of new vertices $v_A$, $v_B$ guarantees that we can recover $G$ from $\SSS(G)$ by a simple interpretation. 
The technical part of the proof is establishing that $\SSS(G)$ comes from a fixed class $\D$ of graphs of tree rank $2$ which depends only on $\C$.


\paragraph*{Organisation of the paper}
After preliminaries in the next section, we introduce vertex rankings in Section~\ref{sec:ranking}, and then we relate them to graph classes of bounded tree rank and show how to compute them in FPT runtime. In Section~\ref{sec:roselemma} we prove a lemma about the behaviour of the $k$-near-twin relation in graphs excluding arbitrarily large half-graphs; a crucial tool for the last section. In Section~\ref{sec:intereps_bd_trr} we prove our main results -- characterization and a sparsification algorithm for graph classes interpretable in classes of tree rank $2$.


\section{Preliminaries}
\label{sec:prelims}

\noindent\textbf{Graph theory.} We use $[k]$ to denote the set $\{1,\ldots, k\}$.
We use mostly standard graph theoretic notation. Let $G$ be a graph. We write $N_r^G(v)$ for the \emph{closed} $r$-neighborhood of a vertex $v$, that is, $N_r^G(v)$ is set of all vertices that are reachable from $v$ by a path with at most $r$ edges, including $v$. The \emph{distance} between vertices $u$ and $v$, denoted by $dist_G(u,v)$, is the minimum number of edges in a path between $u$ and $v$. The \emph{radius} of a graph is smallest integer $r$ so that there exists a vertex that has distance at most $r$ from every other vertex. Given a graph $G$ and a set $X \subseteq V(G)$, we write $G[X]$ for the subgraph of $G$ induced on $X$, and $G-X$ for the subgraph of $G$ induced on $V(G)\setminus X$.

By a \emph{tree} we mean a connected acyclic graph with a specified root vertex. The \emph{depth} (respectively, \emph{height}) of a tree $T$ is the maximum number of edges (respectively, vertices) on any leaf-to-root path in $T$. 

Let $G$ be a graph, and let $A$ and $B$ be subsets of $V(G)$ with $A \cap B = \emptyset$ or $A =B$. By \emph{flipping the edges between  $A$ and $B$}, we mean removing all edges $uv$ in $G$ with $u \in A$ and $v  \in B$, and adding all new edges of the form $uv$ where $u \in A$, $v \in B$, $u\not=v$, and $uv \not\in E(G)$.
For $k \in \N$ and a graph $G$, a \emph{$k$-flip} of $G$ is any graph that can be obtained from $G$ by considering a partition $\F$ of $V(G)$ with $|\F| \le k$, and flipping the edges between some pairs of parts of $\F$ (that is, for each pair of parts $A$ and $B$ of $\F$, we may choose whether or not to flip the edges between $A$ and $B$). We say that a graph class $\C$ is a \emph{perturbation} of a graph class $\D$ if there exists $k$ such that every $G \in \C$ is a $k$-flip of some $H \in \D$.

Let $S$ be a set of vertices of a graph $G$. Let $\F_S$ be the partition of $V(G)$ such that each $v \in S$ is in its own part and all vertices in $V(G) \setminus S$ are partitioned according to their adjacency to $S$ (so vertices with the same neighbors in $S$ are in the same part). An \emph{$S$-flip} of $G$ is any graph $G'$ obtained by flipping the edges between some pairs of parts of $\F_S$.

\vspace{1em}

\noindent\textbf{Shallow topological minors and bounded tree rank.}
Let $H$ be a graph. An \emph{${\le} r$-subdivision} of $H$ is any graph that can be obtained from $H$ by replacing each edge $uv$ of $H$ by a path with endpoints $u$ and $v$ and with at most $r$ internal vertices (so that all of the paths are internally disjoint). We call the original vertices of $H$ the \emph{principal} vertices of the ${\le} r$-subdivision. We say that $H$ is an \emph{$r$-shallow topological minor}\footnote{We note that this definition differs slightly from the standard definition, which says that $H$ is a $r$-shallow minor of $G$ if there is a subgraph of $G$ isomorphic to a graph obtained from $H$ by subdividing its edges at most $2r$ times.} of a graph $G$ if $G$ contains a subgraph that is isomorphic to an ${\le}r$-subdivision of $H$. 

\begin{definition}
We define $T_{d,m}$ to be the tree of depth $d$ in which every non-leaf vertex has exactly $m$ children. 
\end{definition}
It is easily seen that the definition of graph classes of bounded tree rank given in Definition~\ref{def:tree-rank} is equivalent to the following:
\begin{definition}
\label{def:trr}
A graph class $\C$ has tree rank at most $d$ if for every $r\in \mathbb{N}$ there exists $m \in \mathbb{N}$ such that no $G \in \C$ contains $T_{d,m}$ as an $r$-shallow topological minor.
\end{definition}

\noindent\textbf{Strong coloring numbers, admissibility, and bounded expansion.}
Let $G$ be a graph, and let $\le$ be an order on its vertex set. Fix a number $r\in\N$.
For two vertices $v$ and $w$ of $G$, we say that 
$w$ is \emph{strongly $r$-reachable\footnote{We remark that many authors use the order $\le$ in the opposite direction in the definition of strong reachability, that is, they require $w \le v$ and that the path from $v$ to $w$ goes through vertices larger than $v$ in $\le$. The definition we chose will be convenient later on.}} from $v$ (with respect to $\le$) if $w\ge v$ and there is a path from $v$ to $w$ of length at most $r$ in $G$ such that all vertices on this path apart from $v$ and $w$ are smaller than $v$ in $\leq$. The \emph{strong $r$-coloring number} of $G$, denoted by $\scol_r(G)$, is the minimum over all orderings $\le$ of $V(G)$, of the maximum number of vertices that are strongly $r$-reachable from a single vertex $v$ of $G$.

Similarly as above, let $G$ be a graph, and let $\le$ be an order on its vertex set. Fix a number $r\in\N$.
The \emph{$r$-backconnectivity} of a vertex $v$ of $G$ is the maximum number of paths of length at most $r$ in $G$ that start in $v$, end at vertices $w \ge v$, and are vertex-disjoint except for their common endpoint $v$.
The \emph{$r$-admissibility} of $G$, denoted by $\adm_r(G)$, is the minimum over all orderings $\le$ of $V(G)$, of the maximum $r$-backconnectivity of a vertex $v$ of $G$.

Classes of graphs of \emph{bounded expansion} were introduced by Ne\v{s}et\v{r}il and Ossona de Mendez as one of the key notions of their general theory of sparsity~\cite{nevsetvril2012sparsity}. 
We do not provide the original definition of bounded expansion; instead we use the following characterizations.

\begin{theorem}[\cite{zhu2009colouring}]
\label{thm:scol}
    A class $\C$ of graphs has \emph{bounded expansion} if and only if there exists a function $f: \N \to \N$ such that for every $G \in \C$ and $r \in \mathbb{N}$, we have $ \scol_r(G) \le f(r)$.
\end{theorem}
\begin{theorem}[\cite{dvovrak2013constant}]
\label{thm:adm}
    A class $\C$ of graphs has \emph{bounded expansion} if and only if there exists a function $f: \N \to \N$ such that for every $G \in \C$ and $r \in \mathbb{N}$, we have $\adm_r(G) \le f(r)$.
\end{theorem}

\vspace{1em}

\noindent\textbf{Logic, interpretations and locality.} We assume familiarity with  first-order logic and basic notions related to it, such as signatures, quantifier rank, and so on. We model graphs as a structure with one binary irreflexive symmetric
relation $E$. We work with colored graphs, and we model colors as unary predicates.

Interpretations are logic-based transformations that allow us to create new structures from old ones.  
A (simple) \emph{interpretation} $I = (\psi,\delta)$ consists of two formulas $\psi(x,y)$ and $\delta(x)$. When applied to a graph $G$, an interpretation defines a new graph $I(G)$ with
$V(I(G)) = \{ v \in V(G)~:~G \models \delta(v)\}$ and $E(I(G)) = \{uv~:~u,v \in V(I(G)), u\neq v, \text{ and }G \models \psi(u,v)\}$. (Here we assume that $\psi(x,y)$ is symmetric and irreflexive, that is, for all $G$ and $u,v \in V(G)$ we have $G \models \psi(u,v)$ if and only if $G \models \psi(v,u)$, and also for each $v \in V(G)$ we have that $G \not\models \psi(v,v)$ so that the resulting graph is undirected and does not contain loops.) 

For a less general version of interpretation that uses only one formula $\psi(x,y)$, we sometimes use the notation $\psi(G)$ to denote the graph on the same vertex set as $G$ and with $E(\psi(G)):=\{uv~:~u \neq v \textrm{ and }G \models \psi(u,v)\}$. For a graph class $\C$, we say that $\C$ is \emph{interpretable} in a graph class $\D$ if there exists an interpretation $I$ such that $\C \subseteq I(\D)$.

A crucial role in this paper will be played by {interpretations of bounded range}. We say that a formula $\psi(x,y)$ is of \emph{range} $b$ if for all graphs $G$ and all $u,v \in V(G)$ we have that $dist_G(u,v) > b$ implies $G \not\models \psi(u,v)$. This means that if the formula $\psi$ is used in an interpretation, it will not create edges between vertices that were at distance more than $b$ in the original graph. We say that an interpretation $I = (\psi,\delta)$ is of \emph{bounded range} if there exists $b$ such that $\psi$ is of range $b$.

When working with graphs, we often mark vertices of a graph $G$ with (new) unary predicates and call the resulting graph $\wh{G}$. Formally, this means that we are extending the signature of $G$, and that the formulas that we later evaluate on $\wh{G}$ are assumed to be over this new signature. To make the exposition more streamlined, we always do this implicitly.

Locality-based methods are commonly used in relation with first-order logic; probably the most commonly used such result is Gaifman's locality theorem~\cite{gaifman1982local}.
We do not need the full statement of Gaifman's theorem, only the following lemma, which is its simple corollary.
\begin{lemma}
\label{lem:local}
For every formula $\psi(x,y)$, there exist an integer $r$ and a formula $\psi'$ with the following property: Every graph $G$ can be equipped with two unary predicates to obtain $\wh{G}$ such that for any $u,v \in V(G)$ and any $S \subseteq V(G)$ that contains $N_r^{G}(u) \cup N^{G}_r(v)$, we have 
$ G \models \psi(u,v) \Longleftrightarrow \wh{G}[S] \models \psi'(u,v)$.  
\end{lemma}

For completeness we explain how Lemma~\ref{lem:local} follows from Gaifman's theorem. This theorem tells us that for a given formula $\psi(x,y)$ there exists another formula $\alpha(x,y)$ and sentences $\tau_1,\ldots, \tau_k$ such that $\psi(x,y)$ can be written as a boolean combination of $\alpha(x,y), \tau_1,\ldots, \tau_k$. Importantly, the formula $\alpha(x,y)$ has the property that there exists $r$ such that for any $G$ and any $u,v  \in V(G)$ we have $G \models \alpha(u,v)$ if and only if $G[N_r^{G}(u) \cup N^{G}_r(v)] \models \alpha(u,v)$. For any $G$ we can evaluate all sentences $\tau_1,\ldots, \tau_k$ on $G$, and then the boolean combination of $\alpha(x,y), \tau_1,\ldots, \tau_k$ reduces to one of four possibilities on $G$ -- we have that $\psi(x,y)$ is equivalent one of the following: $\alpha(x,y)$, $\lnot\alpha(x,y)$, $\top$ or $\bot$. We can encode these four options with two bits of information, and so we mark all vertices of $G$ with two unary predicates (all vertices in the same way) accordingly. Finally, we define formula $\psi'(x,y)$ to be the formula which first checks the unary predicates on vertex $x$ to determine which of the four possibilities $\phi(x,y)$ is equivalent to on $G$, and based on this `outputs' the value of $\alpha(x,y)$, $\lnot\alpha(x,y)$, $\top$ or $\bot$. To finish the argument, we argue that evaluating $\alpha(x,y)$ on $ \widehat{G}[S]$ gives the same answer as evaluating $\alpha(x,y)$ on $G$ for any $u,v,S$ such that  $N_r^{G}(u) \cup N^{G}_r(v) \subseteq S$. This follows from the properties of $\alpha$ (which hold for any graph, including $\widehat{G}[S]$) which guarantee that $ \widehat{G}[S] \models \alpha(u,v) \Longleftrightarrow G[N_r^{G}(u) \cup N^{G}_r(v)] \models \alpha(u,v) \Longleftrightarrow G \models \alpha(u,v)$.

\section{Vertex rankings in sparse graphs}
\label{sec:ranking}

In this section we introduce a way to assign, for given parameters $r$ and $m$, to every vertex of a graph its \emph{$(r,m)$-rank}. Intuitively, the $(r,m)$-rank of a vertex measures how complicated the $r$-neighborhood of $v$ is. Our definition is algorithmic and is given in Section~\ref{sec:rank_alg} together with the proof that the $(r,m)$--rank can be computed in FPT runtime with respect to the parameters $r$ and $m$. Then we prove Theorems~\ref{thm:rankings=treerank}
and~\ref{thm:rankings=be} relating graph classes for which suitable rankings of vertices exist to graph classes of bounded tree rank (Section~\ref{sec:treerank}) and bounded expansion (Section~\ref{sec:be}).

\subsection{The ranking algorithm}
\label{sec:rank_alg}
We now describe the ranking algorithm that (based on the parameters $r$ and $m$) for every graph $G$ assigns to every vertex of $G$ either a positive integer or $\infty$.

Let $G$ be a graph and $r,m$ be positive integers. The $(r,m)$-ranking algorithm works as follows. Initially, each vertex is assigned rank $\infty$.
Then the algorithm proceeds in rounds for $i=1,2,3,\ldots$ as follows. In the $i$-th round, the algorithm considers all vertices of rank $\infty$ (these are the vertices that have not received a finite rank in rounds $1,\ldots, i-1$), and for each such vertex $v$ it checks (in parallel) the following condition: Does there exist a set $S \subseteq V(G) \setminus \{v\}$ of size at most $m$ such that $N_r^{G - S}(v) \setminus \{v\}$ contains only vertices of finite rank? If yes, then $v$ receives rank $i$, otherwise it keeps rank $\infty$. (Notice that in the first round the algorithm assigns rank 1 to vertices of degree at most $m$.) The algorithm stops when all vertices obtain a finite rank or when in some round no new vertex receives a rank. Thus, in any case, the algorithm stops after at most $|V(G)|$ rounds.

\begin{definition}
Let $G$ be a graph. The \emph{$(r,m)$-rank} of a vertex $v \in V(G)$ is the element of $\N \cup \{\infty\}$ assigned to $v$ by the ranking algorithm.
The \emph{$(r,m)$-rank} of $G$ is the maximum $(r,m)$-rank of any vertex of $G$.   
\end{definition}

We remark that the definition of $(r,m)$-ranking of vertices in a graph $G$ can be phrased without any explicit mention of the ranking algorithm: All vertices of degree at most $m$ are assigned rank $1$, and for every vertex $v$ of degree more than $m$ we define the $(r,m)$-rank of $v$ to be the minimum number $k$ such that there exists a set $S$ of vertices so that $N_r^{G-S}(v) \setminus \{v\}$ contains only vertices of smaller rank; or $\infty$ if such $k$ does not exist. This is easily seen to be equivalent to the algorithmic definition given above, but we prefer to stick with the algorithmic perspective in what follows.

\subsubsection{Algorithmic considerations}

Notice that the straightforward implementation of the ranking algorithm runs in time $\mathcal{O}(n^{m+4})$: we have at most $n$ rounds, in each round we consider at most $n$ vertices, and for each vertex we can in time $\mathcal{O}(n^{m+2})$ try removing all sets $S$ of size $m$ in time $\mathcal{O}(n^m)$ and check whether $G-S$ satisfies the desired condition in time $\mathcal{O}(n^2)$. Thus, we trivially get an XP algorithm. However, we can replace the subroutine which checks whether there exists a set $S$ with $|S| \le m$ with desired properties 
by a simple branch-and-bound procedure with FPT runtime.

\begin{lemma}
\label{lem:routine}
There is an algorithm with runtime $\mathcal{O}(r^m\cdot n^2)$ which correctly decides the following problem: Given $G$, $v \in V(G)$, $A \subseteq V(G)\setminus \{v\}$ and $r, m \in \N$ as input, decide whether there exist a set of set $S \subseteq V(G) \setminus \{v\}$ with $|S| \le m$ such that $N_r^{G - S}(v) \cap A = \emptyset$.
\end{lemma}
\begin{proof}
The algorithm checks whether there is a path of length at most $r$ from $v$ to any vertex in $A$. If no such path exists, the algorithm outputs YES. If such a path $P$ exists and $m = 0$, the algorithm outputs NO. If $m > 0$ then at least one of the vertices on the path $P$ has to be in the solution $S$, and so we can branch on all vertices $u \in V(P)\setminus \{v\}$ (there are at most $r$ of them) and call the algorithm with input $G - \{u\}, v, A \setminus \{u\}$, $r$, $m-1$. Then the algorithm returns YES if for at least one such $u$ the recursive call returned YES, and returns NO otherwise.

We thus get an algorithm with branching bounded by $r$ and depth bounded by $m$, and finding the path from $v$ to $A$ takes time $\mathcal{O}(n^2)$, so the claimed runtime follows.
\end{proof}

As an immediate consequence of the lemma we get the following.
\begin{theorem}
\label{thm:ranking-compute}
The $(r,m)$-ranking of any graph $G$ can be computed in time $\mathcal{O}(r^m\cdot n^4)$.  
\end{theorem}

\subsection{Vertex rankings and bounded tree rank}
\label{sec:treerank}
In this section we prove Theorem~\ref{thm:rankings=treerank}. The proof is split into Lemmas~\ref{lem:treerank1} and~\ref{lem:treerank2}, which correspond to the two directions of the theorem.

\begin{lemma}
\label{lem:treerank1}
Let $G$ be a graph that contains $T_{d,m+1}$ as an $r$-shallow topological minor. Then $G$ has $(r,m)$-rank more than $d$.
\end{lemma}
\begin{proof}
Let $T$ be a subgraph of $G$ that corresponds to an $\le r$-subdivision of $T_{d,m+1}$. We prove that the principal vertices of $T$ of height $i$ (where the height is measured in $T_{d,m+1}$, and we think of leaves as having height $0$) have $(r,m)$-rank at least $i+1$ in~$G$. For $i=0$, the leaves of $T$ clearly have rank at least $1$. For $i>1$, let $v$ be a vertex of $T$ corresponding to a vertex of height $i$ in $T_{d,m+1}$. Then there are $m+1$ internally disjoint paths connecting $v$ to the corresponding $m+1$ children that have rank at least $i$. Therefore we cannot disconnect $v$ from all of these children by deleting $m$ vertices other than $v$. Thus $v$ has $(r,m)$-rank at least $i+1$, which completes the proof.
\end{proof}

\begin{lemma}
\label{lem:treerank2}
Fix any $d,r \in \mathbb{N}$. For every $m$ there exists $m'=m'(d,r,m)$ such that every graph $G$ with $(r,m')$-rank more than $d$ contains $T_{d,m}$ as an $r$-shallow  topological minor.   
\end{lemma}
\begin{proof} 
We will prove the lemma by induction on $d$. Actually, we will prove a slightly stronger statement -- that for a suitably defined $m'$ we have that every vertex of $(r,m')$-rank more than $d$ is the root of an $\le r$-subdivision of $T_{d,m}$ in $G$.

For $d=1$ we can set $m' = m-1$. Then if there exists a vertex of $(r,m')$-rank more than $1$ in $G$, this vertex has at least $m$ neighbors, and thus is the root of a $T_{1,m}$ subgraph in $G$. For $d > 1$ we proceed as follows. 
For a given $m$, we wish to define a suitable $m'$. First, let $W_{d,m,r}$ denote the number of vertices of the graph obtained from $T_{d-1,m}$ by subdividing each edge $r$ times. For convenience, set $M=m\cdot W_{d,m,r}+r\cdot m+m$. Let $m''$ be the number obtained by the inductive assumption applied to $d-1$, $r$, and $M$. Then we define 
$m':= \max\{m'', r\cdot m\}$.
Let $v$ be a vertex of $(r,m')$-rank more than $d$ in $G$. Then for every set $S \subseteq V(G)\setminus \{v\}$ of size at most $m'$, the set $N_r^{G-S}(v)\setminus \{v\}$ contains at least one vertex of $(r,m')$-rank at least $d$.

We claim that there exist paths $P_1, P_2, \ldots, P_{m}$ of length at most $r$ that are disjoint other than at $v$, and so that each path $P_i$ joins $v$ to a vertex $u_i \neq v$ that has $(r,m')$-rank at least $d$. We find the paths greedily by adding one path at a time. Suppose that so far we have found the paths $P_1, \ldots, P_j$ for some $j<{m}$. Let $S$ be the set of all vertices $u \neq v$ that are in any of the paths $P_1, \ldots, P_j$. This set $S$ contains at most $r$ vertices from each of the paths, and thus $S$ has size at most $r(m-1) \leq {m'}$. Thus the set $N_r^{G-S}(v)\setminus \{v\}$ contains at least one vertex of $(r,m')$-rank at least $d$, and we can add another path $P_{j+1}$ to our collection. This proves the claim. We note that this argument amounts to a Menger-type result about short paths, and this type of result has previously appeared in~\cite{mengerShort}.

Since $m' \geq m''$, the $(r, m'')$-rank of each vertex of $G$ is at least its $(r, m')$-rank. So by the inductive assumption applied to $d-1$, $r$, and $M$, each vertex $u_i$ is the root of an $\le r$-subdivision of $T_{d-1,M}$. Let us call this $\le r$-subdivision rooted at $u_i$ by $T_i$.

We now greedily build up the desired $r$-shallow topological minor of $T_{d,m}$ rooted at $v$ as follows. Suppose that for some $j<m$, we have found that $T_1, \ldots, T_j$ contain, respectively, subgraphs $T_1', \ldots, T_j'$ so that \begin{itemize}
\item for every $i \in \{1,\ldots, j\}$, the subgraph $T_i'$ of $T_i$ is an $\leq r$-subdivision of $T_{d-1,m}$ that is rooted at $u_i$, 
\item the subgraphs $T_1', \ldots, T_j'$ are pairwise vertex-disjoint, and
\item for every $i \in \{1,\ldots, j\}$, the vertex $u_i$ is the only vertex that is in both $T_i'$ and in any of the paths $P_1, \ldots, P_m$. 
\end{itemize}
\noindent We begin this greedy process with $j=0$. Now suppose that it holds for some $j<m$. We will prove it for $j+1$. This will complete the proof of Lemma~\ref{lem:treerank2}.

So, we need to find a subgraph $T_{j+1}'$ of $T_{j+1}$ so that $T_{j+1}'$ is an $\leq r$-subdivision of $T_{d-1,m}$ that is rooted at $u_{j+1}$, and the only vertex in common between $T_{j+1}'$ and any of $T_1', \ldots, T_j'$ or $P_1, \ldots, P_m$ is $u_{j+1}$. Notice that each of the trees $T_1', \ldots, T_j'$ has at most $W_{d,m,r}$ vertices by the definition of $W_{d,m,r}$. Moreover, each of the paths $P_1, \ldots, P_m$ contributes at most $r$ additional vertices. So, since $j <m$, the number of vertices in any of $T_1', \ldots, T_j'$ or $P_1, \ldots, P_m$ is at most $m\cdot W_{d,m,r}+r\cdot m$. Recall from the definition that $M$ is this quantity plus $m$. So we can just delete the branches of $T_{j+1}$ in which any of the vertices of $T_1', \ldots, T_j'$ or $P_1, \ldots, P_m$, other than $u_{j+1}$, occur. In this manner we arrive at the desired subgraph $T_{j+1}'$.
\end{proof}



\subsection{Vertex rankings and bounded expansion}
\label{sec:be}

In this section we prove Theorem~\ref{thm:rankings=be}, characterizing classes of bounded expansion using vertex rankings. The theorem follows from the two lemmas below combined with Theorems~\ref{thm:scol} and~\ref{thm:adm}.

\begin{lemma}
\label{lem:be:1}
    Let $G$ be a graph with $\scol_r(G) = m$. Then $G$ has finite $(r,m-1)$-rank.
\end{lemma}

We stress that in Lemma~\ref{lem:be:1} we are not bounding the maximum value of $(r,m-1)$-rank of any vertex of $G$, but merely claiming that the $(r,m-1)$-ranking algorithm will assign to each vertex of $G$ a finite value. This value, however, can be arbitrarily large (and in particular it is not bounded in terms of $r$ and $m$). This can be seen already on the class of trees -- the class of all trees has bounded strong coloring numbers (one can set $m:=r+1$), but on the complete $m$-ary tree of depth $d$ the $(r,m-1)$-ranking algorithm will need $d$ rounds to finish.

\begin{proof}[Proof of Lemma~\ref{lem:be:1}]
Assume for contradiction that there exists a vertex of $G$ of $(r,m-1)$-rank $\infty$.
Fix an order $\le$ on $V(G)$ that certifies that $\scol_r(G) = m$.
Let $v$ be the smallest vertex of $G$ with respect to $\le$ that has $(r,m-1)$-rank $\infty$. 
Let $A$ be the set of all vertices below $v$ in $\le$, and let $i$ be the maximum $(r,m-1)$-rank of any vertex in $A$. We claim that in round $(i+1)$ of the ranking algorithm vertex $v$ received rank (meaning it has rank $i+1$), which is a contradiction with our assumption on $v$. To see this, let $S$ be the set of all vertices that are strongly $r$-reachable from $v$ in $G$. Note that every path of length at most $r$ from $v$ to a vertex $w \ge v$ must contain a vertex in $S$. It follows that $N_r^{G-S}(v) \setminus \{v\}$ contains only vertices of finite rank, a contradiction.
\end{proof}

Lemma~\ref{lem:be:1} guarantees that the ranking algorithm succeeds on graphs with bounded strong coloring numbers. The next lemma tells us that the ranking algorithm computes a good admissibility ordering for an input graph $G$.

\begin{lemma}
\label{lem:be:2}
Let $r,m \in \N$ with $r\ge 1$.
Let $G$ be a graph with finite $(r,m)$-rank. Then $\adm_r(G) \le m$.
\end{lemma}

\begin{proof}
Let $\le$ be an ordering of $V(G)$ so that if $v \le w$, then the $(r,m)$-rank of $v$ is at most the $(r,m)$-rank of $w$. Thus we are ordering vertices of $G$ by their rank, breaking ties arbitrarily.
We claim that this ordering certifies that $\adm_r(G) \le m$.
Let $v$ be any vertex of $G$, and let $i$ be its $(r,m)$-rank. Let $S$ be the set of vertices certifying that the $(r,m)$-rank of $v$ is $i$. That is, $S$ is a subset of $V(G)\setminus \{v\}$ of size at most $m$ so that $N_r^{G-S}(v)\setminus \{v\}$ contains only vertices of $(r,m)$-rank strictly less than $i$.

Since any vertex $w \ge v$ has $(r,m)$-rank at least $i$, any path from $v$ that ends in a vertex larger than $v$ with respect to the ordering must contain a vertex of $S$.
Therefore, if we consider a collection $P_1,\ldots, P_\ell$ of paths that maximizes the $r$-backconnectivity of $v$, then each $P_i$ has to intersect $S$. Since all of these paths are disjoint except for their common end $v$, we get that $\adm_r(G) \le m$, as desired.
\end{proof}

\section{A lemma about edge-stable graphs}
\label{sec:roselemma}
In this section we state and prove a lemma (Lemma~\ref{lem:Rose_lemma}) about the behaviour of the so-called $k$-near-twin relation on graphs that do not contain large half-graphs. This lemma will be crucial in the next sections and may be of independent interest in the theory of (monadically) stable graphs~\cite{shelah1971stability,shelah1972combinatorial} (see also~\cite{dreier2023firstorder} for the latest important developments in this area).

First we need a few definitions.
A crucial role will be played by the notion of {$k$-near-twins}.

\begin{definition}
Let $G$ be a graph, and let $k \in \N$.
We say that two vertices are \emph{$k$-near-twins} if $|N^G(v) \Delta N^G(u)| \le k$.
\end{definition}

\begin{definition}
Let $G$ be a graph, and let $k \in \N$.
The \emph{$k$-near-twin graph of $G$}, denoted by $NT_k(G)$, is the graph with vertex set $V(G)$ in which there is an edge between two vertices $u$ and $v$ if they are $k$-near-twins in $G$.
\end{definition}

\begin{definition}
A \emph{half-graph of order $t$} is a graph with vertex set $\{u_1,\ldots,u_t, w_1,\ldots, w_t\}$ such that there is an edge between $u_i$ and $w_j$ if and only if $i \le j$.
\end{definition}

\begin{definition}
We say that a graph $G$ contains a half-graph of order $t$ as a \emph{semi-induced subgraph} if there are distinct vertices $u_1,\ldots,u_t, w_1,\ldots, w_t$ in $G$ such that there is an edge between $u_i$ and $w_j$ if and only if $i \le j$.   
\end{definition}

Note that in the definition we do not say anything about edges between the vertices within the set $\{w_1,\ldots, w_t\}$ or edges between the vertices within the set $\{u_1,\ldots, u_t\}$; the edges within these sets can be arbitrary.

Our main lemma in this section is the following.

\begin{lemma}
\label{lem:Rose_lemma}
There is a function $h:\N^2 \to \N$
so that for any $k,t \in \N$, if $G$ is a graph with no half-graph of order $t$ as a semi-induced subgraph, and $u$ and $v$ are vertices in the same component of the $k$-near-twin graph of $G$, then $u$ and $v$ are $h(k,t)$-near-twins in $G$.
\end{lemma} 

Lemma~\ref{lem:Rose_lemma} will follow easily from the following technical lemma. This lemma finds distinct vertices $w_1, w_2, \ldots, w_t$ along with corresponding sets of vertices $X_1, X_2, \ldots, X_t$. The first two properties of the lemma imply that each set $X_i$ is contained in the common neighborhood of $w_i, w_{i+1}, \ldots, w_t$. The third property implies that there are ``many'' vertices in $X_i$ that are not a neighbor of any of $w_1, w_2, \ldots, w_{i-1}$. These properties will help us prove Lemma~\ref{lem:Rose_lemma}.

\begin{lemma}
\label{lem:rose_technical}
Set $g(c,k,1) := c$ for all $c,k \in \mathbb{N}$, and by induction on $t$ set $g(c,k,t):=g(c,k,t-1)\cdot(t-1) + k +c$. Let $c,k,t \in \N$ be arbitrary with $t \ge 1$. 
 Let $P:=v_1 v_2\ldots v_m$ be a path in $NT_k(G)$, and suppose that $S$ is a subset of $N^G(v_m)\setminus N^G(v_1)$ of size at least $g(c,k,t)$.
Then there are distinct vertices $w_1,\ldots, w_t$ of $P$ and distinct subsets $X_1,\ldots, X_t$ of $V(G)$ such that for each $i \in [t]$ we have

\begin{enumerate}
    \item $X_i \subseteq N^G(w_i) \cap S$,
    \item $X_{i-1} \subseteq X_i$ (here we set $X_0 = \emptyset$), and
    \item $|X_{i}| \ge |X_i \cap N^G(w_{i-1})| +  \ldots + |X_i \cap N^G(w_{1})| +c$.
\end{enumerate} 
\end{lemma}

\begin{proof}
We proceed by induction on $t$. For $t=1$, we can set $w_1= v_m$ and $X_1 = S$ and the conditions above are easily seen to be satisfied.

For $t > 1$ we set $w_t := v_m$. Let $q$ be the smallest number such that $|S \cap N^G(v_q)| \ge g(k,t-1)$. Note that $q$ exists since $|S \cap N^G(v_m)| \ge g(c,k,t) \ge g(c,k, t-1)$. Moreover, $q$ is strictly less than $m$ since $v_{m-1}$ and $v_m$ are $k$-near-twins in $G$ and $g(c,k,t) \ge g(c,k, t-1)+k$.

Set $X_t:=S$. 
To show the existence of the remaining vertices $w_1,\ldots,w_{t-1}$ and sets $X_1, \ldots, X_{t-1}$, we will use the induction hypothesis with $t-1$ applied to the path $P' := v_1\ldots v_q$  and the set $S' := S \cap N^G(v_q)$ (this is justified since $S' \subseteq N^G(v_q)$, the size of $S'$ is as required, and $S' \subseteq S$, so $S'$ is disjoint from $N^G(v_1)$). This yields vertices $w_1,\ldots, w_{t-1}$ and sets $X_1,\ldots, X_{t-1}$ with properties 1-3 above satisfied for $P'$ and $S'$. We now verify that $w_1, \ldots, w_t$ together with $X_1,\ldots, X_t$ satisfy properties 1-3 for $P$ and $S$. Let $i \in [t]$ be arbitrary.
\begin{enumerate}
    \item For $i=t$ we have $X_t = S$ and $w_t = v_m$, and so  $X_t = S = N^G(w_t) \cap S$, as desired. For $i < t$ we have from the induction hypothesis that $X_i \subseteq N^G(w_i) \cap S'$, which is a subset of $N^G(w_i) \cap S$ because $S' \subseteq S$.

    \item For $i=t$ we nave $X_t = S$ and since $X_{t-1} \subseteq S'$ (by the induction hypothesis) and $S' \subseteq S$, we have $X_{t-1} \subseteq X_t$. For $i < t$ the property holds since it holds for all $X_1,\ldots, X_{t-1}$.

    \item For $i=t$ we have $|X_t| = |S| \ge g(c,k,t)$. Note that by the choice of $q$ we have that for every vertex $v_j$ from $\{v_1,\ldots, v_{q-1}\}$ it holds that $|S \cap N^G(v_j)| < g(c,k,t-1)$. For $v_q$ we have that $|S \cap N^G(v_q)| < g(c,k,t-1) + k$, because $q$ is a $k$-near-twin of $v_{q-1}$ and $|S \cap N^G(v_{q-1})| < g(c,k,t-1)$ and so there cannot be more than $k$ vertices in $S$ for which their adjacency differs. Since all vertices $w_1,\ldots, w_{t-1}$ are all from $\{v_1,\ldots, v_q\}$ and $X_t = S$, we have that $|X_t \cap N^G(w_j)| < g(c, k,t-1)$ for every $w_j$ with $j < t$, except for possibly one $w_j$ which may be equal to $v_q$. In this case we would have $|X_t \cap N^G(w_j)| < g(c,k,t-1) + k$. In total, we have 
    $$ \left(\sum_{j = 1}^{t-1} |X_t \cap N^G(w_j)|\right)+c \le (t-1)g(c,k,t-1)+ k +c = g(c,k,t) \le |X_t|,$$
    as desired. For $i < t$ the result follows from the induction hypothesis.
\end{enumerate}
This completes the proof of Lemma~\ref{lem:rose_technical}.
\end{proof}
   
\begin{proof}[Proof of Lemma~\ref{lem:Rose_lemma}]
Set $c := t+1$, and set $h(k,t) :=2g(c,k,t)$, where $g$ is the function from Lemma~\ref{lem:rose_technical}. Assume for a contradiction that there are vertices $u,v$ in the same connected component in the $k$-near-twin graph of $G$ that are not $h(k,t)$ twins. We will construct a half-graph of order $t$ as a semi-induced subgraph of $G$.

Since $u$ and $v$ are not $h(k,t)$-near-twins, we have $|N^G(v) \setminus N^G(u)| \ge g(c,k,t)$ or $|N^G(u) \setminus N^G(v)| \ge g(c,k,t)$; without loss of generality assume that the first option holds. Since $u$ and $v$ are in the same connected component of $NT_k(G)$, they are connected by a path $P=v_1\ldots v_m$ in $NT_k(G)$ where $v_1 =u$ and $v_m = v$. Set $S:=N^G(v) \setminus N^G(u)$. Then the assumptions of Lemma~\ref{lem:rose_technical} are satisfied, and we have in $G$ vertices $w_1,\ldots, w_t$ and sets $X_1,\ldots, X_t$ with properties 1-3 from the lemma. We construct a semi-induced half-graph of order $t$ in $G$, where we put vertices $w_1,\ldots, w_t$ on one side and for the other side we pick vertices $u_1,\ldots, u_t$ with $u_i \in X_i$ as follows. 

We pick $u_1$ from $X_1\setminus \{w_1, w_2, \ldots, w_t\}$ arbitrarily; then we have $w_1u_1 \in E(G)$ since $X_1 \subseteq N^G(w_1)$. For $i>1$, assuming that $u_1,\ldots, u_{i-1}$ have been selected, we proceed as follows. By property 3 of Lemma~\ref{lem:rose_technical}, there is a subset $M_i$ of $X_i$ containing at least $t+1$ vertices in $X_i$ that are not neighbors of any of $w_1 \ldots w_{i-1}$. We pick as $u_i$ any vertex from $M_i \setminus \{w_1,w_2,\ldots, w_{t}\}$. This vertex $u_i$ is not adjacent to $w_j$ for any $j < i$ since it was chosen from $M_i$. (So this vertex also must be different from $u_1, u_2, \ldots, u_{i-1}$.) Moreover, we have that  $w_i$ is adjacent to all $u_j$ with $j \le i$, because every such vertex $u_j$ is in $X_j$ (by construction from previous steps), and $X_j 
\subseteq X_i \subseteq N^G(w_i)\cap S$. Thus, we have $u_iw_j \in E(G)$ if and only if $i \le j$, as desired.
\end{proof}

\section{Interpretations of graph classes of tree rank 2}
\label{sec:intereps_bd_trr}

In this section we prove Theorems~\ref{thm:characterization} and~\ref{thm:sparsify_alg}.
First we will establish some properties of classes of tree rank at most $2$ (Section~\ref{sec:tr2}). After this, in Section~\ref{sec:core}, we give a characterization of graph classes interpretable in graph classes of tree rank at most $2$ by interpretations of \emph{bounded range}. This section contains the technical core of our results -- the key result is Theorem~\ref{thm:sparsify_bounded_range}, proof of which is split into Lemmas~\ref{lem:forward} and~\ref{lem:main}.  The case of general interpretations (Theorem~\ref{thm:characterization}) is presented in Section~\ref{sec:full_interps}; it will easily follow from results in Section~\ref{sec:core}. Finally, we prove our main algorithmic result, Theorem~\ref{thm:sparsify_alg}, in Section~\ref{sec:algo}.

\subsection{Classes of tree rank at most 2}
\label{sec:tr2}
We start with a characterization of graph classes of tree rank $2$.
\begin{definition}
\label{def:labd}
A class of graphs $\C$ has \emph{locally almost\footnote{We remark that the adjective `almost' is sometimes used differently in context of structural graph parameters. For example, for a graph class $\C$, having almost bounded flipwidth (see~\cite{flipwidth}) means that for every $\epsilon$ there exists $c$ such that every $G \in \C$ has flipwidth at most $cn^\epsilon$. Our usage of the adjective `almost' is different -- it refers to having a bounded number of exceptions.} bounded degree} if there exist functions $f,d: \N \to \N$ such that for every $r \in \N$, every $G \in \C$, and every $v \in V(G)$, the set $N^G_r(v)$ contains at most $f(r)$ vertices with degree larger than $d(r)$ in $G$.
\end{definition}

\begin{lemma}
\label{lem:trr2=lbd}
A class $\C$ of graphs has tree rank at most $2$ if and only if has locally almost bounded degree.
\end{lemma}

In the proof of the lemma we will use the following simple result from~\cite[Lemma~13]{gajarsky2024elementary}.
\begin{lemma}
\label{lem:many_reachable}
Let $r,t \in \N$, let $H$ be a graph of radius at most $r$, and let $S\subseteq V(H)$. If $|S| \ge t^r+1$, then there exists $u \in V(H)$ such that there are $t$ vertices in $S$ that are distinct from $u$ and can be reached from $u$ by internally vertex-disjoint paths of length at most $r$.
\end{lemma}

\begin{proof}[Proof of Lemma~\ref{lem:trr2=lbd}]
By Theorem~\ref{thm:rankings=treerank}, it suffices to prove that a class $\C$ is of locally almost bounded degree if and only if for every $r \in \mathbb{N}$ there exists $m \in \mathbb{N}$ such that every graph in $\C$ has $(r,m)$-rank at most $2$.

First, assume that $\C$ is of locally almost bounded degree with respect to functions $f$ and $d$. Let $r \in \mathbb{N}$, and set $m:=\max\{f(r), d(r)\}$. We claim  that the $(r,m)$-ranking algorithm will assign the number $1$ or $2$ to each $v \in V(G)$. In the first round, all vertices of degree at most $d(r) \le m$ will get rank $1$. In the second round, we know that each $N_r^G(v)$ contains at most $f(r)$ vertices of degree larger than $d(r)$. Thus we can delete a set $S \subseteq V(G)\setminus \{v\}$ of size at most $f(r)$ so that $N_r^{G-S}(v)\setminus \{v\}$ contains only vertices of rank $1$. The result follows.

For the other direction, suppose that for every $r \in \mathbb{N}$ there exists $m \in \mathbb{N}$ such that every graph in $\C$ has $(r,m)$-rank at most $2$. Let $r \in \mathbb{N}$, and set $f(r):=(m+1)^r$ and $d(r) :=m$. Let $v \in V(G)$ be arbitrary, and let $S$ be the set of vertices with $(r,m)$-rank exactly $2$ in $N_r^G(v)$ (these are the vertices of degree more than $m$). Going for a contradiction, we may assume that $|S|>f(r)$, since otherwise we are done. Thus, by Lemma~\ref{lem:many_reachable} applied to $r$, $t:=m+1$ and $S$ in the subgraph $H$ of $G$ induced on $N_r^G(v)$, there exists a vertex $u \in N_r^G(v)$ such that there are $m+1$ vertices in $S$ that are distinct from $u$ and can be reached from $u$ by internally vertex-disjoint paths of length at most $r$. As all of the vertices in $S$ have $(r,m)$-rank $2$, this shows that the $(r,m)$-rank of $u$ is more than $2$, which is a contradiction. Thus $\C$ is of locally almost bounded degree with functions $f$ and $d$, as desired.
\end{proof}

\noindent \textbf{Algorithmic considerations.}
For our algorithms we will need to assume that the functions which are used in the definitions of our graph classes are efficiently computable. In particular, we will need to be able to efficiently test whether a given graph $G$ comes from a fixed graph class of locally almost bounded degree given by functions $d$ and $f$. This can be done under fairly relaxed conditions on functions $d$ and $f$, which we now describe. The key observation is that for any graph $G$ we need to check the conditions imposed by functions $f$ and $d$ from Definition~\ref{def:labd} only for values of $r$ with $r \le n$ (where $n=|V(G)|$), and that checking the conditions is trivial whenever $f(r)>n$ or $d(r)> n$.

With this in mind, we say that a function $h: \N \to \N$ is \emph{nice} if there exists a an algorithm which inputs two numbers $r$ and $n$ represented in binary with $r \le n$ and in time $poly(n)$ either correctly answers that $h(r) > n$  or 
outputs $h(r)$ (which is upper bounded by $n$).

Note that a function which is very fast growing and  complicated to compute (with respect to its input which has length $\lceil \log(r) \rceil$) can still be nice. This is because it may be easy to check that $h(r) > n$, in which case no further computation is required, and if $h(r) \le n$, we know that $n$ is much larger than $\lceil \log(r) \rceil$, and then we have $poly(n)$ time at our disposal to compute $h(r)$. It is easy to verify that functions such as $2^r$, $\text{tow}_\ell(r)$ (tower of twos of height $\ell$ with $r$ on top), $\text{tow}_r(2)$ (tower of twos of height $r$) and also $r^r$ are nice functions. 

Coming back to graph classes of tree rank $2$, we will use the following definition to specify graph classes for which we can prove Theorem~\ref{thm:sparsify_alg}. 
\begin{definition}
We say that a class $\C$ of graphs of tree rank $2$ is \emph{efficiently bounded} if there exists a nice function $g: \N \to \N$ such that for every $r\in \mathbb{N}$ there exists $m \in \mathbb{N}$ such that no $G \in \C$ contains $T_{2,m}$ as an $r$-shallow topological minor.   
\end{definition}

By inspecting the proof of Lemma~\ref{lem:trr2=lbd} one easily checks that any efficiently bounded class $\C$ of tree rank $2$ is a class of locally almost bounded degree for which the there exist functions $f$ and $d$ which ceritfy this. 
If we moreover assume that  $\C$ is \emph{maximum} class of locally bounded degree with respect to $f$ and $g$ (meaning it contains every $G$ that satisfies conditions given by $f$ and $d$ for all $r \le |V(G)|$), then we can efficiently test for any $G$ whether $G \in \C$. This is easily achieved by testing the conditions from Definition~\ref{def:labd} for all values $r$ up to $|V(G)|$. Since the functions $f$ and $d$ are nice, these values are either too large (if $f(r)>n$ or $d(r)>n$ which can be efficiently checked), or efficiently computable. This yields the following lemma.
\begin{lemma}
\label{lem:check_lbdg}
Let $\C$ be a class of graphs with locally almost bounded degree given by nice functions $f,d: \N \to \N$. Assume that $\C$ is maximum such class with respect to $f$ and $d$. Then there is a polynomial time algorithm that determines whether $G \in \C$.
\end{lemma}

In subsequent sections, it would be desirable to show at various places that some function is nice (or at least can be upper bounded by some nice function). We will not spell out these arguments explicitly; we just note here that 
one can often combine two nice functions to obtain another nice function.
In particular, one easily checks that if $f$ and $g$ are functions such that for all $r$ we have $f(r) \ge r$ and $g(r) \ge r$, then also $h:= g \circ f$ is nice with $h(r) \ge r$. This because for given $r$, $n$ with $r \le n$ we can check whether $f(r) > n$, and if yes then we know that $h(r) = g(f(n)) > n$. On the other hand if $f(r) \le n$, then the input condition for function $g$ is satisfied, and we can just check whether $g(f(r)) > n$ or compute the value of $g(f(r))$ efficiently.


\subsection{The case of interpretations of bounded range}
\label{sec:core}

We now proceed by giving a characterization of graph classes that can be obtained from graph classes of tree rank at most $2$ by an interpretation of bounded range (Theorem~\ref{thm:sparsify_bounded_range}).

In what follows we will use the notions of $k$-near-twins and near-twin graphs introduced in Section~\ref{sec:roselemma}.

\begin{definition}
\label{def:near-covered}
We say that a graph $G$ is \emph{$(k,m)$-near-covered} if
for every set $S\subseteq V(G)$ such that the elements of $S$ are mutually not $k$-near-twins we have $|S| \le m$. 
\end{definition}
We remark that the definition of $(k,m)$-near-covered graphs  introduced in~\cite{bd_deg_interp2} was slightly different -- there it was required that there exists a set of at most $m$ vertices such that every vertex $v$ of $G$ is a $k$-near-twin of some $w \in S$. One can easily check that the two definitions are functionally equivalent (i.e. up to changing $k$ and $m$ in one definition to $k'$ and $m'$ in the other definition). 
We use Definition~\ref{def:near-covered} because it will be more convenient in our arguments.

\begin{definition}
    A graph class $\C$ is \emph{locally almost near-covered} if there exist functions $k,m: \N \to \N$ such that for every $r\in \mathbb{N}$, every $G \in C$, and every $v \in V(G)$, the subgraph of $G$ induced on $N_r^G(v)$ is $(k(r),m(r))$-near-covered.
\end{definition}

\begin{theorem}
\label{thm:sparsify_bounded_range}
    A class $\C$ is a interpretable from a class of locally almost bounded degree by a bounded range interpretation if and only if $\C$ is locally almost near-covered. 
\end{theorem}

The forward direction of Theorem~\ref{thm:sparsify_bounded_range} is handled by the following lemma, which will be proven by combining the locality-based Lemma~\ref{lem:local} with the characterization of graph classes interpretable in graph classes of bounded degree in terms of near-covered graphs given in~\cite{bd_deg_interp2}.

\begin{lemma}
\label{lem:forward}
Let $I$ be a bounded range interpretation, and let $\C$ be a class of locally almost bounded degree. Then $I(\C)$ is locally almost near-covered.
\end{lemma}

In the proof we will use the following lemma.
\begin{lemma}[\cite{bd_deg_interp2}]
\label{lem:bd_deg_interp}
Let $\D$ be a graph class interpretable in a class of graphs of bounded degree. Then there exist $k$ and $m$ such that  every $G \in \D$ is $(k,m)$-near-covered.
\end{lemma}

\begin{proof}[Proof of Lemma~\ref{lem:forward}]
First we describe an auxiliary construction. Fix $d\in \N$.
Let $H$ be a graph and let $v_1,\ldots v_{k}$ be the vertices of $H$ of degree more than $d$.  We define the graph $\mathcal{K}(H)$ to be the graph obtained from $H$ in which we
\begin{itemize}
    \item isolate every vertex $v_i$ (that is, remove all edges incident to $v_i$),
    \item mark $v_i$ with a unary predicate $L_i$, and
    \item mark the neighbors of $v_i$ (in $H$) with a unary predicate $P_i$.
\end{itemize}
It is easily seen that for every $k$, if $H$ has at most $k$ vertices of degree more than $d$, then there exists a quantifier-free formula $\delta_k$ that recovers $H$ from $\mathcal{K}(H)$, i.e. we have $H = \delta_k(\mathcal{K}(H))$.

Now, let $f$ and $d$ be the functions certifying that the class $\C$ is of locally almost bounded degree, and let $b$ be the integer certifying that $I$ is of bounded range. Fix an arbitrary integer $r$. Let $G \in \C$ and $v \in V(G)$. Our goal is to show that the subgraph of $G$ induced on $N_r^G(v)$ is $(k(r),m(r))$-near-covered for suitable functions $k$ and $m$ of $r$.

Let $H \in \C$ be such that $G = I(H)$. Let $r'$ and $\psi'$ be the number and formula obtained from Lemma~\ref{lem:local} applied to $\psi$. Thus $H$ can be equipped with two unary predicates to obtain $\widehat{H}$ such that for any $u,w \in V(H)$ and any $S \subseteq V(G)$ that contains $N_{r'}^H(u) \cup N_{r'}^H(w)$, 
$$H \models \psi(u,w) \Longleftrightarrow  \widehat{H}[S] \models \psi'(u,w). $$ 

We will apply this fact to a particular set $S$, which we now define. First of all, since $\psi$ is of range $b$, we know that $N_r^G(v)$ is a subset of $N_{br}^H(v)$. Also, for every $u \in N_{br}^H(v)$ we have that $N_{r'}^H(u) \subseteq N_{br+r'}^H(v)$. We set $S:=N_{br+r'}^H(v)$. Thus, for any $u,w \in N_{br}^H(v)$, we have that $H \models \psi(u,w) \Longleftrightarrow  \wh{H}[S] \models \psi'(u,w)$. Since we have $uw \in E(G) \Leftrightarrow H \models \psi(u,w)$, 
we get that $G[N_r^G(v)]$
is an induced subgraph of $\psi'(\wh{H}[S])$. So the proof of the lemma will be finished if we show that $\psi'(\wh{H}[S])$ is $(k(r),m(r))$-near-covered for suitably chosen $k$ and $m$.

Since $H$ is from a graph class of locally almost bounded degree and the graph $\wh{H}[S]$ has radius at most $br+r'$, we have that $\wh{H}[S]$ contains at most $f(br+r')$ vertices of degree more than $d(br+r')$. Thus, if we set $k:=f(br+r')$, then the formula $\delta_k$ defined at the beginning of the proof recovers $\wh{H}[S]$ from $\mathcal{K}(\wh{H}[S])$. The graph $\mathcal{K}(\wh{H}[S])$ is from a graph class of degree at most $d(br+r')$, and $\psi'(\wh{H}[S]) = \psi'(\delta_k(\mathcal{K}(H_v))$. Thus the graph $\psi'(\wh{H}[S])$ is obtained by applying an interpretation to a graph from a class of degree at most $d(br+r')$. The lemma now follows by applying Lemma~\ref{lem:bd_deg_interp}.
\end{proof}


The rest of this section is dedicated to proving the backward direction of Theorem~\ref{thm:sparsify_bounded_range}



\begin{lemma}
\label{lem:main}
Let $\C$ be a locally almost near-covered graph class. Then $\C$ is interpretable in a class $\D$ of locally almost bounded degree by an interpretation $I$ of bounded range. 

Moreover, there is a polynomial algorithm that to every graph $G$ computes a graph $\SSS(G)$ so that if $G \in \C$, then $\SSS(G) \in \D$ and $G = I(\SSS(G))$.
\end{lemma}

We will need the following simple lemma showing that locally almost near-covered graph classes do not contain arbitrarily large half-graphs.

\begin{lemma}
\label{lem:no_ladder}
Let $\C$ be a locally almost near-covered graph class with respect to functions $k$ and $m$, and set $t:=m(2)\cdot k(2) + m(2)+1$. Then no graph in $\C$ contains a half-graph of order $t$ as a semi-induced subgraph.
\end{lemma}
\begin{proof}
Assume for contradiction that such a graph $G \in \C$ exists.
Let $u_1,\ldots, u_t,w_1,\ldots, w_t$ be the vertices of a half-graph of order $t$ that $G$ contains as a semi-induced subgraph. Then all of these vertices $u_1,\ldots, u_t,w_1,\ldots, w_t$ are in $N_2^G(u_1)$. Also, for any $i,j \in \{1,2,\ldots, t\}$ with $i-j > k(2)$, we know that $u_i$ and $u_j$ are not $k(2)$-near twins, since their adjacency differs in vertices  $w_{j},w_{j+1}, \ldots, w_{i-1}$. Therefore, the vertices $u_1, u_{k(2)+2}, u_{2k(2)+3},\ldots u_{m(2)\cdot k(2) + m(2)+1}$ form a set of $m(2)+1$ pairwise not $k(2)$-near-twins, a contradiction with $\C$ being locally almost near-covered.
\end{proof}

In the proof of Lemma~\ref{lem:main} we will also  use the following lemma taken from~\cite[Corollary~5.3]{bd_deg_interp2}. We note that in that paper, an extra assumption was made that amounts to saying that no vertex in $A$ is a $k$-near-twin of any vertex in $B$. However, that assumption was not used in the proof given in~\cite{bd_deg_interp2}, and so same proof works to prove the following lemma.

\begin{lemma}
\label{lem:almost_all}
Let $G$ be a graph and let $A$ and $B$ be two subsets of $V(G)$ that are either disjoint or have $A=B$ and such that $|A| \ge 5k+1$, $|B| \ge 5k+1$, all vertices in $A$ are pairwise $k$-near-twins, and all vertices in $B$ are pairwise $k$-near-twins. Then either
\begin{enumerate}
\item every vertex of $A$ is adjacent to at most $2k$ vertices of $B$ and every vertex of $B$ is adjacent to at most $2k$ vertices of $A$, or
\item every vertex of $A$ is adjacent to all but at most $2k$ vertices of $B$ and every vertex of $B$ is adjacent to all but at most $2k$ vertices of $A$
\end{enumerate}
\end{lemma}

We will also use the following immediate corollary of Lemma~\ref{lem:almost_all}.

\begin{corollary}
\label{cor:flip_bd_deg}
Let $G$ be a graph, and let $A$ and $B$ be two subsets of $V(G)$ that satisfy the assumptions of Lemma~\ref{lem:almost_all}. If there exists a vertex $v\in A$ with $|N^G(v) \cap B| > 2k$, then in the graph $G'$ obtained from $G$ by flipping the edges between $A$ and $B$, every vertex $u \in A$ has $|N^{G'}(u) \cap B| \le 2k$.
\end{corollary}

Finally, we will use the following Ramsey-type result, which is routine to prove.

\begin{lemma}
\label{lem:RamseyStar}
There exists functions $R,D:\mathbb{N}^3 \rightarrow \mathbb{N}$ so that for all $k, m, s \in \mathbb{N}$, the following holds for every graph $G$ with no subgraph isomorphic to the complete bipartite graph $K_{s,s}$. If $X \subseteq V(G)$ is such that $|X| \geq R(k,m,s)$ and every vertex in $X$ has degree at least $D(k,m,s)$, then there exists $Y \subseteq X$ such that $|Y| \geq m$ and every vertex $y \in Y$ has at least $k$ neighbors in $V(G)\setminus Y$ that are not adjacent to any other $y' \in Y\setminus \{y\}$.
\end{lemma}

\begin{proof}[Proof of Lemma~\ref{lem:main}]
Let $\C$ be a locally almost near-covered graph class, and let $k$ and $m$ be the functions certifying this. We will describe a procedure $\mathcal{S}$ that to every $G \in \C$ assigns a graph $\mathcal{S}(G)$ such that $G = I(\mathcal{S}(G))$ for a fixed interpretation $I$ of bounded range. This procedure will also return some graph $\mathcal{S}(G)$ even when $G$ is not in $\C$, however in this case we make no guarantees about the graph $\mathcal{S}(G)$. We will then show that the graph class $\D:=\{\mathcal{S}(G)~:~G \in \C\}$ is of locally almost bounded degree.
Since by construction we will have $\C \subseteq I(\D)$, this will finish the proof.

Let $G \in \C$ be arbitrary. Let $h:=h(k(3), t)$, where $h$ is the function from Lemma~\ref{lem:Rose_lemma} and $t:=m(2)\cdot k(2) + m(2)+1$ is the integer from Lemma~\ref{lem:no_ladder}. Thus every pair of vertices in the same component of the $k(3)$-near-twin graph of $G$ are $h$-near-twins in $G$. We define a partition $\F$ of $V(G)$ by putting two vertices into the same part if they are in the same connected component of the $k(3)$-near-twin graph of $G$ (this creates a partition since being in the same connected component is an equivalence relation). We call parts $A,B \in \F$ (possibly with $A=B$) \emph{mutually heavy} if $|A| \geq 5h+1$, $|B| \geq 5h+1$, and there exists a vertex $v \in B$ with $|N^G(v) \cap A| > 2h$. (Note that in this case, by Lemma~\ref{lem:almost_all}, there also exists a vertex $u \in A$ with $|N^G(u) \cap B| > 2h$.) Similarly, we call a part $A \in \F$ \emph{heavy} if it is mutually heavy with any part $B \in \F$ (possibly with $B=A$).\\

\noindent \textbf{Creating the sparse  graph $\boldsymbol{\SSS(G)}$ from $\boldsymbol{G}$.}
We can now describe the sparse graph $\mathcal{S}(G)$ associated to $G$.
We start from $G$ and proceed as follows: 
\begin{enumerate}
    \item For any heavy part $A \in \F$, we introduce a new vertex $v_A$, make it adjacent to all vertices in $A$, and mark it with a unary predicate $R$. 
    \item  For any mutually heavy parts $A,B \in \F$, we flip between $A$ and $B$. If $A=B$, then we mark $v_A$ with a unary predicate $F$. If $A \neq B$, then we put an edge between $v_A$ and $v_B$.
\end{enumerate} 
All other adjacencies that were not part of any flip remain as in $G$. Note that we can define this graph $\SSS(G)$ even when $G$ is not in $\C$. Also note that by Corollary~\ref{cor:flip_bd_deg}, for all mutually heavy parts $A,B \in \F$ and every vertex $v \in A$, we have that $|N^{\SSS(G)}(v) \cap B| \leq 2h$. We will use this important property later on.\\

\noindent \textbf{Algorithmic considerations.} It is easily seen that the construction of $\SSS(G)$ can be done in polynomial time from $G$.\\


\noindent \textbf{Recovering $\boldsymbol{{G}}$ from $\boldsymbol{{S(G)}}$  by an interpretation.}
We now show that there is an interpretation $I$ such that $G = I(\mathcal{S}(G))$. Let $u$ and $v$ be two vertices from $V(G)$. In the construction of $\mathcal{S}(G)$, the adjacency between two vertices only changed in the second part of the construction. Thus, the formula $\psi(x,y)$ will complement the adjacency between distinct vertices $x$ and $y$ if
\begin{itemize}
    \item $x$ and $y$ have the same (unique) neighbor $w$ marked with the predicate $R$ and this $w$ is also marked with predicate $F$, or
    \item $x$ and $y$ each have a different neighbor marked with the predicate $R$ and these neighbors are adjacent.
\end{itemize}
In all other cases we have $\psi(x,y)=E(x,y)$. The conditions above are easily expressed by a first-order formula. Note that $\psi(x,y)$ only keeps the existing edges or creates edges between vertices at distance at most $3$ and therefore $\psi$ is of range $3$. To define the vertex set of $G$ from the graph $\mathcal{S}(G)$, the formula $\delta(x)$ just keeps the vertices that are not marked with $R$ (the original vertices of $G$). \\

\noindent \textbf{Showing that the class $\boldsymbol{
\D := \{\mathcal{S}(G) : G \in \C\}}$ is of locally almost bounded degree.} 
The rest of the proof is devoted to showing that $\D$ is of locally almost bounded degree. First we define an auxiliary graph $G/\F$, which we call the \emph{quotient graph}. This graph has vertex-set $\F$, and two vertices $A,B \in \F$ are adjacent in $G/\F$ if there exist vertices $u \in A$ and $v \in B$ such that $uv \in E(G)$. We also note that if parts $A,B \in \F$ are mutually heavy, then $B \in N^{G/\F}(A)$. This fact will be used throughout the proof.

We have already defined which parts of $\F$ are \emph{heavy}. Now, let us say that a part $A$ of $\F$ is \emph{light} if there exists a vertex $v \in A$ of degree at most $h$ in $G$. Note that no part is both light and heavy, but there may be some parts that are neither.

We now establish some basic properties of the graph $G/\F$ and the partition $\F$ of $V(G)$. First we prove two claims that show that vertices in a heavy part (or two mutually heavy parts) are close to each other in $G$. In fact we prove something slightly stronger.

\begin{claim}
\label{claim:dist_two}
For any part $A \in \F$ that is not light and any $u,v \in A$, we have $\dist_G(u,v) \le 2$.
\end{claim}
\begin{claimproof}
 By Lemma~\ref{lem:Rose_lemma}, the vertices $u$ and $v$ are $h$-near-twins in $G$, which means that $|N^G(u) \Delta N^G(v)| \le h$. Since $A$ is not light, the vertex $u$ has degree larger than $h$ in $G$, and so at least one neighbor of $u$ in $G$ is also adjacent to $v$.
\end{claimproof}

\begin{claim}
\label{claim:dist_three}
For any distinct mutually heavy parts $A,B \in \F$ and any $u\in A$ and $v \in B$, we have $\dist_G(u,v) \le 3$.
\end{claim}
\begin{claimproof}
  By the definition of mutually heavy parts, the vertex $v$ has some neighbor $u'$ that is in $A$. Since $u$ and $u'$ each have at most $2h$ non-neighbors in $B$ and $|B|> 4h$, they have a common neighbor. It follows that $\dist_G(u,v) \leq 3$.
\end{claimproof}

Next we show that among the non-light parts, the quotient graph has bounded degree. Similarly to before, we actually show the following slightly stronger claim.

\begin{claim}
\label{claim:Gamma_bd_deg}
For any part $A \in \F$ that is not light, all but at most $m(3)$ neighbors of $A$ in $G/\F$ are light.
\end{claim}

\begin{claimproof}
 For convenience set $t:=m(3)+1$. Assume for contradiction that $A$ has neighbors $B_1,B_2, \ldots, B_{t}$ in $G/\F$ that are not light. Then there exist $a_1,a_2,\ldots, a_{t} \in A$ and $b_1 \in B_1, b_2 \in B_2, \ldots, b_{t} \in B_t$ such that for each $i \in \{1,\ldots t\}$ we have that $a_ib_i \in E(G)$. 
By Claim~\ref{claim:dist_two} we know that the distance between any $a_i$ and $a_j$ is at most $2$ in $G$. Thus all vertices $b_1, b_2,\ldots, b_t$ are at distance at most $3$ from $a_1$. Since for any $i,j$ with $i\not=j$ we have that $b_i$ is not a $k(3)$-near-twin of $b_j$ (because they come from different parts of $\F$), this contradicts the fact that $\C$ is locally almost near-covered with respect to $k$ and $m$.
\end{claimproof}

Next comes a key claim that says that sets of small radius in $\SSS(G)$ essentially come from sets of small radius in $G$. Of course we are only able to show something like this for pairs of vertices that are actually in $V(G)$.

\begin{claim}
\label{claim:distPath}
For any pair of vertices $u$ and $v$ of $G$, we have $\dist_{G}(u,v) \leq 3\cdot\dist_{\SSS(G)}(u,v)$.
\end{claim}

\begin{claimproof}
Consider a shortest path $P$ between $u$ and $v$ in $\SSS(G)$. By breaking up this path into smaller parts, we may assume that no internal vertex of $P$ is also a vertex of $G$. First suppose that $P$ has just one edge $uv$. If $uv \in E(G)$, then we are done. Otherwise, there exist mutually heavy parts $A,B \in \F$ so that $u \in A$ and $v \in B$. If $A=B$ then we are done by Claim~\ref{claim:dist_two}, and if $A \neq B$ then we are done by Claim~\ref{claim:dist_three}. So we may assume that $P$ has more than one edge. This case also follows by combining Claims~\ref{claim:dist_two} and~\ref{claim:dist_three}.
\end{claimproof}

We are ready to take care of the high-degree vertices that are not vertices of $G$. We will take care of the other case (of high degree vertices that are in $V(G)$) later on.

\begin{claim}
\label{claim:outsideOfG}
For any vertex $v$ of $\SSS(G)$ and any $r \in \mathbb{N}$, there are less than $8(m(r')+1)(m(3)+1)^2$ vertices in $N_r^{\SSS(G)}(v) \setminus V(G)$ that have degree larger than $k(r')+2h$ in the graph $\SSS(G)$, where we set $r':=6r+24$ for convenience.
\end{claim}

\begin{claimproof}
Going for a contradiction, suppose otherwise. We set $t := 8(m(r')+1)(m(3)+1)^2$ for convenience. Thus, there are distinct heavy parts $A_1, A_2, \ldots, A_t \in \F$ so that all of the vertices $v_{A_1}, v_{A_2}, \ldots, v_{A_t}$ are in $N_r^{\SSS(G)}(v)$ and have degree larger than $k(r')+2h$. For each $i \in \{1,2,\ldots, t\}$, fix an arbitrary arbitrary part $B_i$ such that $A_i$ and $B_i$ are mutually heavy (it is possible that $B_i = A_i$), and fix arbitrary vertices $a_i \in A_i$ and $b_i \in B_i$. 

We claim that all of $b_1, b_2, \ldots, b_t$ are in $N_{r'}^{G}(b_1)$. First we will show that all of $b_1, b_2, \ldots, b_t$ are in $N_{2r+8}^{\SSS(G)}(b_1)$; then the desired statement follows from Claim~\ref{claim:distPath} since $r' = 3(2r+8)$. So, to see this, first note that every vertex $v_{A_i}$ has distance at most $2r$ from $v_{A_1}$ in $\SSS(G)$ since both of these vertices are in $N_r^{\SSS(G)}(v)$. Then $a_i$ has distance $1$ from $v_{A_i}$. Finally, by Claims~\ref{claim:dist_two} and~\ref{claim:dist_three}, the vertex $b_i$ has distance at most $3$ from $a_i$, which shows what we want.

Now let us define an auxiliary graph $\wh{G}$ where each vertex of $\wh{G}$ is a tuple of the form $(A_i, B_i)$, and two distinct vertices $(A_i, B_i)$ and $(A_j, B_j)$ are adjacent if $A_i \cup B_i$ has non-empty intersection with $A_j \cup B_j$. So this graph $\wh{G}$ has exactly $t$ vertices. Moreover, it has maximum degree less than $4(m(3)+1)$; for each vertex $(A_i, B_i)$, there are at most $2m(3)+1$ ways of choosing a tuple $(A, B)$ of mutually heavy parts of $\F$ so that either $A=A_i$ or $B=A_i$, and likewise with $B_i$ in place of $A_i$. This is using Claim~\ref{claim:Gamma_bd_deg} which implies that each vertex of $G/\F$ is mutually heavy with at most $m(3)$ other parts of $\F$ (and the extra $+1$ comes from the fact that a vertex may be mutually heavy with itself).

Thus, since $\wh{G}$ has $t$ vertices and maximum degree less than $4(m(3)+1)$, it has an independent set of size at least $t/(4(m(3)+1))=2(m(r')+1)(m(3)+1)$. Let $I \subseteq \{1,2,\ldots, t\}$ be a set of this size such that the vertices $((A_i, B_i): i \in I)$ form an independent set in $\wh{G}$. By the same type of argument about the maximum degree of an auxiliary graph, we can find a set $I' \subseteq I$ of size $|I|/2(m(3)+1)=m(r')+1$ so that for all distinct $i,j \in I'$, we have that $B_i$ is not adjacent to $A_j$ in the quotient graph $G/\F$. 

Finally, consider the two vertices $b_i$ and $b_j$ for distinct $i,j \in I'$. The vertex $b_i$ has no neighbors in $A_j$ in $G$. However, since $A_j$ and $B_j$ are mutually heavy, the vertex $b_j$ does have at least $|A_j|-2h$ neighbors in $A_j$. Since $v_{A_j}$ has degree larger than $k(r')+2h$ in $\SSS(G)$, the set $A_j$ has size larger than $k(r')+2h$. It follows that $b_i$ and $b_j$ are not $k(r')$-near-twins in $G$. Since all of $b_1, b_2, \ldots, b_t$ are in $N_{r'}^{G}(b_1)$, this is a contradiction to the fact that $G$ is locally almost near-covered by functions $k$ and $m$.
\end{claimproof}

We need a couple more helpful claims before we can take care of the high-degree vertices of $\SSS(G)$ that are in $V(G)$.

\begin{claim}
\label{claim:samePartTwins}
For any part $A \in \F$, every pair of vertices $u,v \in A$ are $h$-near-twins in~$\SSS(G)$.
\end{claim}

\begin{claimproof}
The vertices $u$ and $v$ have the same adjacencies to the vertices in $V(\SSS(G))\setminus V(G)$. Moreover, any time we flipped between two sets, each of those sets either contained both of $u$ and $v$, or neither of them. Since $u$ and $v$ are $h$-near-twins in $G$, the claim follows.
\end{claimproof}

Next we show that the graph $\SSS(G)$ is ``weakly sparse'', that is, it forbids a small complete bipartite graph as a subgraph.

\begin{claim}
\label{claim:weaklySparse}
Set $s:=5h\cdot m(3)+1$. Then the graph $\SSS(G)$ does not have a subgraph that is isomorphic to the complete bipartite graph $K_{s,s}$.
\end{claim}
\begin{claimproof}
Suppose towards a contradiction that there exist disjoint sets $X,Y \subseteq V(\SSS(G))$ of size $s$ so that every vertex in $X$ is adjacent to every vertex in $Y$. Since $s>2h$, no vertex in $X$ or $Y$ is in a part of $\F$ that is light. 

Next, we claim that there exist $X' \subseteq X$ and $Y' \subseteq Y$ each of size $5h+1$ so that $X'$ is contained in a single part $A\in \F$ and $Y'$ is contained in a single part $B \in \F$. By symmetry, we just need to show this for $Y$. So, let $x \in X$. By Claim~\ref{claim:Gamma_bd_deg} applied to the part of $\F$ that contains $x$, we find that $Y$ is contained in the union of at most $m(3)$ parts of $\F$. So indeed, the desired set $Y'$ exists by the pigeonhole principle.

Now, the parts $A$ and $B$ that contain $X'$ and $Y'$ have size at least $5h+1$, and there is a vertex in $A$ that has more than $2h$ neighbors in $B$. However, this contradicts Corollary~\ref{cor:flip_bd_deg} and our choice of when to flip between two parts $A$ and $B$.
\end{claimproof}

For the rest of the proof, fix $s$ as in Claim~\ref{claim:weaklySparse}. We also fix the functions $R,D:\mathbb{N}^3\to \mathbb{N}$ from Lemma~\ref{lem:RamseyStar}, the Ramsey lemma. In the final key claim, we bound the number of vertices in a small-radius neighborhood in $\SSS(G)$ that have high degree and are vertices of~$G$. 

\begin{claim}
\label{claim:highDegG}
For any vertex $v$ of $\SSS(G)$ and any $r \in \mathbb{N}$, there are at most $R(k', m(6r)+1, s)$ vertices in $N_r^{\SSS(G)}(v) \cap V(G)$ that have degree larger than $D(k', m(6r)+1, s)$ in $\SSS(G)$, where we set $k':=(5h+1)\cdot m(3)+k(6r)+1$ for convenience.
\end{claim}

\begin{claimproof}
Going for a contradiction, suppose otherwise. Let $X$ be the set of all vertices in $N_r^{\SSS(G)}(v) \cap V(G)$ that have degree larger than $D(k', m(6r)+1, s)$ in $\SSS(G)$. Recall from Claim~\ref{claim:weaklySparse} that $\SSS(G)$ has no subgraph isomorphic to the complete bipartite graph $K_{s,s}$. So by Lemma~\ref{lem:RamseyStar}, there exists $Y \subseteq X$ such that $|Y| \geq m(6r)+1$ and every vertex $y \in Y$ has at least $k'$ neighbors in $\SSS(G)$ that are not in $Y$ and are not adjacent to any other vertex $y' \in Y \setminus \{y\}$. Let $N_y$ denote the set of all such neighbors of $y$.

Now, fix an arbitrary vertex $x\in X$. Note that $Y \subseteq N_{2r}^{\SSS(G)}(x)$ since every pair of vertices in $Y$ is in $N_{r}^{\SSS(G)}(v)$. So by Claim~\ref{claim:distPath}, we have that $Y \subseteq N_{6r}^G(x)$. Thus, since $|Y|>m(6r)$ and $G$ is almost near-covered, there exist two vertices $y,y' \in Y$ that are $k(6r)$-near-twins in $G$. 
Since $y$ and $y'$ both have at least $k' > k(6r)$ `private' neighbors (in $N_y$ and $N_{y'}$, respectively), they are not $k(6r)$-near-twins in $\SSS(G)$, and so at least one of them had its neighborhood changed when creating $\SSS(G)$ from $G$. It follows that at least one of $y$, $y'$ is in a heavy part of $\F$; without loss of generality assume it is $y$ and let $A$ be the set in $\F$ with $y \in A$. Our goal will now be to prove that there exists $B \in \F$ such that parts $A,B$ are mutually heavy in $G$ and $|N^{\SSS(G)}(y) \cap B| > 2h$.  This will finish the proof, as it is a contradiction with the fact that in  $\SSS(G)$ there is no vertex in $A$ with $|N^{\SSS(G)}(y) \cap B| > 2h$, which follows from the construction of $\SSS(G)$ and Corollary~\ref{cor:flip_bd_deg}.

First, note that there cannot be more than $k(6r)$ vertices in $N_y$ that are in light parts of $\F$, because otherwise $y$ and $y'$ could no be $k(6r)$-near-twins in $G$. This is because the adjacencies of vertices from light parts remained unchanged in the construction of $\SSS(G)$ (so $y$ sees the same vertices in light parts in $G$ as in $\SSS(G)$) and no vertex in $N_y$ is a neighbor of $y'$.




Finally, by Claim~\ref{claim:Gamma_bd_deg}, the set $A$ has at most $m(3)$ neighbors in the quotient graph $G/\F$ that are not light. Since $y$ has at most one neighbor in $\SSS(G)$ that is not a vertex of $G$, it follows that there exists $N_y' \subseteq N_y$ of size $(|N_y|-k(6r)-1)/m(3)=5h+1$ so that $N_y'$ is all contained in a single part $B$ of $\F$. Clearly this part $B$ has at least $5h + 1$ vertices, and so by the construction of $\SSS(G)$ vertex $y$ from $A$ cannot have more than $2h$ neighbors in $B$, a contradiction.

\end{claimproof}

By combining Claims~\ref{claim:outsideOfG} and~\ref{claim:highDegG}, we find that the graph $\SSS(G)$ has locally almost bounded degree for suitable functions $f$ and $d$. This completes the proof of Lemma~\ref{lem:main}.

\end{proof}

\vspace{-15pt}
\subsection{The case of full interpretations}
\label{sec:full_interps}

In this section we finish the proof of our main result, Theorem~\ref{thm:characterization}, which states that a class $\C$ of graphs is interpretable in a graph class of tree rank $2$ if and only if $\C$ is a perturbation of a locally almost near-covered graph class.

For the forward direction, it is known (see for example~\cite{blcw, horizons}) that every interpretation can decomposed into two steps in the following sense.

\begin{lemma}
\label{lem:interps}
For every interpretation $I$ there exists an interpretation $I'$ of bounded range and $k \in \N$ such that for every $G$ we have that $I(G)$ is a $k$-flip of $I'(G)$.
\end{lemma}

The forward direction of Theorem~\ref{thm:characterization} then follows immediately from Lemmas~\ref{lem:forward} and~\ref{lem:interps}. The backward direction follows from Lemma~\ref{lem:main}: If $\C$ is a perturbation of a locally almost near-covered graph class, then $\C$ is a perturbation of $I(\D)$ for some graph class $\D$ of locally almost bounded degree. Since perturbations ($k$-flips) can be modelled by interpretations, add unary predicates to graphs from $\D$ and adjust $I$ to $I'$ such that $\C \subseteq I'(\D)$.

\subsection{Proof of Theorem~\ref{thm:sparsify_alg}}
\label{sec:algo}

In this section we prove Theorem~\ref{thm:sparsify_alg} by showing that for every graph class $\C$ interpretable in a graph class of tree rank $2$, there is a polynomial time algorithm that to every $G \in \C$ computes $H \in \D$ such that $G = I(H)$, where $I$ is a fixed interpretation and $\D$ is a fixed class of graphs of locally almost bounded degree (which is the same as tree rank $2$ by Lemma~\ref{lem:trr2=lbd}).

We will rely on the following lemma which was proven in~\cite{blcw}.
\begin{lemma}
\label{lem:incremental}
Let $\D$ be an NIP class of graphs, and let $I$ be an interpretation. Then there exist $s$, $b$, and a formula $\psi(x,y)$ of range $b$ such that for every $G \in I(\D)$ there exist $S\subseteq V(G)$ of size at most $s$, an $S$-flip $G'$ of $G$, and a graph $H \in \D$ such that  $G' = \psi(H)$.

\end{lemma}

We do not define the notion of a graph class being NIP (see for example the relevant sections in~\cite{blcw}), but just note that this notion is very general and it is easily established that classes of tree rank $2$ are NIP, so the lemma can be applied in our setting.

We now proceed with the proof Theorem~\ref{thm:sparsify_alg}.
Since  $\C$ is interpretable in a graph class of tree rank $2$, there exist an interpretation $I$ and a class $\D$ of tree rank at most $2$ such that $\C \subseteq I(\D)$. By applying Lemma~\ref{lem:incremental} to $\D$ and $I$, we obtain numbers $s$ and $b$ and a formula $\psi$ of range $b$. Since $\psi$ is of bounded range, we know by Lemma~\ref{lem:forward} that the class $\psi(\D)$ is locally almost near-covered. By Lemma~\ref{lem:main} applied to $\psi(\D)$, there exists a graph class $\D'$ of locally almost bounded degree, an interpretation $I'$, and a polynomial time algorithm $\mathcal{A}$ that to any $G' \in \psi(\D)$ computes a graph $\SSS(G') \in \D'$ such that $G' = I'(\SSS(G'))$. Moreover, by inspecting the proofs of Lemma~\ref{lem:forward} and Lemma~\ref{lem:main} one can check that the functions $f$ and $d$ which certify that $\D'$ is a class of locally almost bounded degree are nice functions (and we can also assume $\D'$ to be maximum with respect to $f$ and $d$), and so we can use Lemma~\ref{lem:check_lbdg} to check membership in $\D'$.

We now consider the algorithm that for a graph $G \in \C$ does the following:

\begin{enumerate}
    \item Go through all subsets $S$ of $V(G)$ of size at most $s$.
    \item For each such set $S$, go through all possible $S$-flips $G'$ of $G$.
    \item Apply the algorithm $\mathcal{A}$ from Lemma~\ref{lem:main} to $G'$ to obtain a graph $\SSS(G')$.
    \item Check whether $G' = I'(\SSS(G'))$ and use Lemma~\ref{lem:check_lbdg} to check whether $\SSS(G') \in \D'$, if yes, then output $\SSS(G')$.
\end{enumerate}

We now argue the correctness of the algorithm.
By Lemma~\ref{lem:incremental}, at least one $G'$ will be such that $G' = \psi(H)$ for some $H \in \D$. Then, since $G' \in \psi(\D)$, we have (by Lemma~\ref{lem:main}) that the algorithm $\mathcal{A}$ returns a graph $\SSS(G')$ that is in $\D'$. Since for $\SSS(G')$ we have $G'=I'(\SSS(G'))$ and  since from $G'$ one can easily recover $G$ by performing a bounded number of flips, we can adjust the interpretation $I'$ to an interpretation $I''$ such that $G = I''(\SSS(G'))$, as desired. (We add unary predicates to $\SSS(G')$ in order to ``mark'' the sets that we flip between.)

The runtime of the algorithm is easily seen to be polynomial in $|V(G)|$: We have $|V(G)|^s$ possible sets of size $s$ and consequently we have $|V(G)|^s$ iterations, where the number $s$ depends only on the graph class $\C$. In each iteration we invoke the polynomial time algorithm $\mathcal{A}$ from Lemma~\ref{lem:main}.

\section{Conclusions}

We conclude with two open ended questions that may deserve further attention.
\begin{enumerate}
\item Is there a way of defining vertex rankings for dense graphs analogous to our rankings? Ideally, such rankings should be computable in FPT time.
\item While the proof of Lemma~\ref{lem:main} is technical, our construction of the graph $\SSS(G)$ from the graph $G$ is simple:
First we determine which components of the $k$-near-twin graph of $G$ to flip between, and then we use this information to construct $\SSS(G)$. It may be interesting to see under which conditions on $G$ we can claim that the graph $\SSS(G)$ comes from a class of sparse graphs. Also, using our construction recursively may lead to interesting results: If $\SSS(G)$ is not a sparse graph, one may consider $\SSS(\SSS(G))$, and so on.

\end{enumerate}

\bibliography{biblio}

\end{document}